\documentclass[showpacs,twocolumn]{revtex4}

\usepackage{graphicx}
\usepackage{dcolumn}
\usepackage{amsmath}
\usepackage{amssymb}

\begin{document}

\title{Extensive Chaos in the Lorenz-96 Model}

\author{A. Karimi}
\affiliation{Department of Engineering Science and Mechanics, Virginia
Polytechnic and State University, Blacksburg, Virginia 24061}
\author{M.R. Paul}
 \email{mrp@vt.edu}
\affiliation{Department of Mechanical Engineering, Virginia
Polytechnic and State University, Blacksburg, Virginia 24061}


\begin{abstract}
We explore the high-dimensional chaotic dynamics of the Lorenz-96 model by 
computing the variation of the fractal dimension with system parameters. The 
Lorenz-96 model is a continuous in time and discrete in space model first 
proposed by Edward Lorenz to study fundamental issues regarding the 
forecasting of spatially extended chaotic systems such as the atmosphere. First, 
we explore the spatiotemporal chaos limit by increasing the system size while 
holding the magnitude of the external forcing constant. Second, we explore 
the strong driving limit by increasing the external forcing while holding the system 
size fixed. As the system size is increased for small values of the forcing we 
find dynamical states that alternate between periodic and chaotic dynamics. 
The windows of chaos are extensive, on average, with relative deviations 
from extensivity on the order of 20\%. For intermediate values of the 
forcing we find chaotic dynamics for all system sizes past a critical value. The 
fractal dimension exhibits a maximum deviation from extensivity on the order of 5\% for small 
changes in system size and decreases non-monotonically with increasing 
system size.  The length scale describing the deviations from extensivity 
and the natural chaotic length scale are approximately equal in support of the 
suggestion that deviations from extensivity are due to the addition of chaotic 
degrees of freedom as the system size is increased.  As the forcing is increased at 
constant system size the fractal dimension exhibits a power-law dependence.  
The power-law behavior is independent of the system size and quantifies the 
decreasing size of chaotic degrees of freedom with increased forcing which we 
compare with spatial features of the patterns. 
\end{abstract}


\maketitle

\section{Introduction}

A common feature of spatially extended systems that are driven far-from-equilibrium 
is spatiotemporal chaos where the dynamics are aperiodic in space and 
time~\cite{cross:1993}. Important examples include the dynamics of the 
atmosphere and climate~\cite{lorenz:1968}; fluid convection~\cite{cross:1993,bodenschatz:2000}; 
the convection of biological organisms in the oceans~\cite{bees:1997}, the transition 
to chaos in excitable media~\cite{bar:1993}; and the complicated dynamics of 
systems of reacting, advecting and diffusing chemicals~\cite{nugent:2004}.  
Despite intense theoretical and experimental investigation many important 
challenges remain regarding how to characterize, control, and predict the dynamics of 
these large complex systems.

A unifying feature of many of these systems is that the dimension of the attractors describing their 
dynamics in phase space is very large.  This severely limits the use of 
sophisticated techniques of chaotic time-series analysis~\cite{abarbanel:1996}
and of powerful geometry based descriptions of dynamical systems~\cite{wiggins:2003}. 
However, the use of Lyapunov exponents to compute the fractal dimension does 
not suffer from these limitations and provides an important window into  
fundamental aspects of the dynamics of high dimensional chaotic systems. 
Furthermore, the spectrum of Lyapunov exponents are inaccessible to experimental 
measurement using currently available techniques which leaves their numerical 
computation as an important way to gain new insights. A physical understanding 
of the dimension and how it varies with system parameters can provide fundamental 
knowledge regarding the underlying nature of spatiotemporal chaos that can be used 
to guide the development of improved theoretical descriptions for experimentally 
accessible systems. 

A defining feature of chaos is the sensitive dependence of the system dynamics 
on initial conditions and the exponential separation of nearby trajectories in 
phase space~\cite{lorenz:1963}. The rate of separation is quantified by the spectrum 
of Lyapunov exponents $\lambda_i$ where $i=1,\ldots,N$ with $\lambda_i$ arranged 
in descending order and $N$ is the total number of degrees of freedom in the system. 
For partial differential equations, although the phase space is infinite, the value of $N$ 
is expected to be very large but finite corresponding to the dimension of the inertial 
manifold (c.f.~\cite{robinson:1995,yang:2009}). For a system of coupled ordinary 
differential equations, as we study here, the value of $N$ is equal to the number of 
independent variables. The summation of the first $M$ exponents (where $M \le N$) 
describes the growth of an $M$-dimensional ball of initial conditions in $N$-dimensional 
phase space. The exact interpolated number of exponents required for the sum to 
vanish is the dimension of the ball that will neither grow nor shrink and is an 
estimate of the fractal dimension $D_\lambda$ of the strange attractor. 
Using a linear interpolation yields the well known Kaplan-Yorke 
formula
\begin{equation}
 D_\lambda=K+ \sum_{i=1}^K \frac{\lambda_i} {\left| \lambda_{K+1} \right|} 
\end{equation}
where $K$ is the largest integer such that the sum of the first $K$ Lyapunov 
exponents is non-negative~\cite{ott:1993}.  The magnitude of $D_\lambda$ is an approximate  
value of the number of degrees of freedom acting in the system~\cite{ruelle:1985,farmer:1983}.

Ruelle~\cite{ruelle:1982} initially conjectured, and numerical simulation later 
confirmed (c.f.~\cite{ohern:1996,manneville:1985}), that the fractal dimension 
is extensive for large chaotic systems.  More precisely,  $D_\lambda$ increases  
linearly with the volume of the system $L^d$ where $L$ is a characteristic 
length and $d$ is the number of spatially extended dimensions. The extensivity of 
$D_\lambda$ is a consequence of spatial disorder. If two subsystems are 
sufficiently far apart their coupling is weak and they contribute independently and 
additively to the overall dimension~\cite{ruelle:1985,cross:1993}.  Cross and 
Hohenberg~\cite{cross:1993} used this to define a natural chaotic length scale 
as
\begin{equation}
\xi_\delta = \left(\frac{D_\lambda}{L^d} \right)^{-1/d}
\label{eq:chaotic_length}
\end{equation}
where a volume $\xi_\delta^d$ will contain, on average, one degree of freedom.

Using the above arguments, one intriguing possibility is that underlying 
spatiotemporal chaos are structures of volume $\sim \xi_\delta^d$. This raises an 
interesting question: How does the dimension vary for changes in system 
volume that are smaller than $\xi_\delta^d$? One possibility is that 
$D_\lambda$ is proportional to the system size only on average.  In this case, the variation 
of $D_\lambda$ would have a stepwise structure with system size where the 
steps correspond to the addition of new degrees of freedom as the system grows large enough to 
accommodate them. A smoother transition may also be possible where the new 
degrees of freedom are able to stretch or compress to match the system 
size.  In light of this, the variation of $D_\lambda$ for small changes in system 
size can shed important physical insight upon the basic nature and 
composition of spatiotemporal chaos~(c.f.~\cite{cross:1993,fishman:2006,tajima:2002,xi:2000}).

Questions relating to the precise variation of the fractal dimension with system size 
have been explored numerically for several important model equations. 
O'Hern~\emph{et al.} explored the extensive chaos of a 2D coupled-map 
lattice with an Ising-like transition~\cite{ohern:1996}.  The fractal dimension was found to exhibit significant 
deviations from extensivity near the onset of spatiotemporal chaos which rapidly decayed 
with increasing system size (here the number of lattice sites).

Xi~\textit{et al.}~\cite{xi:2000} explored the 1D Nikolaevskii equation and Tajima 
and Greenside~\cite{tajima:2002} explored the 1D Kuramoto-Sivashinsky 
equation.  In both of these studies extensive chaos was found and a systematic 
exploration of the dimension was performed in the extensive 
regime. The fractal dimension remained linear to within the precision of the calculations 
to yield what is now referred to as microextensive chaos.   It is important to note that 
neither study focussed upon the approach to extensive chaos where deviations may 
be more accessible. At the larger system sizes explored it remains a possibility that 
deviations in the fractal dimension were too small to detect.

Fishman and Egolf explored the 1D complex Ginzburg-Landau equation over a range of system 
sizes including the approach to extensive chaos~\cite{fishman:2006}. Significant deviations from 
extensivity were found to occur on a length scale consistent with $\xi_\delta$ which 
was used to suggest the presence of building blocks that compose spatiotemporal chaos~\cite{fishman:2006}.

Computations of the fractal dimension have also been conducted for experimentally 
accessible systems such as Rayleigh-B\'{e}nard convection (the buoyant convection that 
occurs in a shallow fluid layer heated from below). Extensive chaos 
was found in large periodic boxes~\cite{egolf:2000} and in finite cylindrical 
domains~\cite{paul:2007} yielding a natural chaotic length scale $\xi_\delta \approx 2$ 
corresponding approximately to the width of a convection roll pair.  However, 
such calculations are very expensive and it remains unclear if a fluid system 
such as Rayleigh-B\'{e}nard convection is microextensive.

The approach we take is to study the dynamics of a periodic lattice 
of coupled variables (discrete in space and continuous in time) whose 
equations of motion were originally developed by Edward Lorenz in 1996 
with the dynamics of the atmosphere and fluid convection in mind~\cite{lorenz:1996}. 
The model has since become important in the study of forecasting spatiotemporally 
chaotic systems. The model is now known as the Lorenz-96 model and is described 
in detail below.  We compute the spectrum of Lyapunov exponents and fractal dimension 
using standard approaches. The contribution of this paper is a systematic 
and careful study of the variations in the fractal dimension with changes 
in system parameters to gain physical insights into 
high-dimensional chaos. By exploring a simple model we are able to perform 
very long-time simulations, over many initial conditions, and for a broad 
range of system parameters.

\section{The Lorenz-96 Model}

The Lorenz-96 model is a simple model developed by Edward Lorenz 
in 1996 to study difficult questions regarding predictability in weather 
forecasting~\cite{lorenz:1996}. The model is constructed of variables 
that represent the continuous time variation of an atmospheric quantity 
of interest, such as temperature or vorticity, at a discrete location on a periodic lattice 
representing a latitude circle on the earth.  The discrete variables are 
coupled spatially and their equation of motion includes contributions relevant 
to fluid systems including a quadratic nonlinearity, dissipation, and a constant 
external forcing.  However, the equations are phenomenological and can not 
be derived systematically from a more rigorous description.

Despite these simplifications the Lorenz-96 model has emerged as an important and often 
used model system for the testing of new ideas in the atmospheric 
sciences~\cite{lorenz:1998,boffetta:2002,ott:2004} and in the general study of 
spatiotemporal chaos~\cite{pazo:2008}. In our work, we explore the 
Lorenz-96 model as a numerically accessible 
model with phenomenological relevance to fluid systems.  The computational 
cost of a systematic study of the variation of the fractal dimension 
with system parameters is significant. It is anticipated that our exploration 
of a simple model will yield new insights that can be used to guide 
future studies of more complicated systems.

Mathematically, the Lorenz-96 model is a linear lattice of $N$ variables  
where the dynamics of the $k$th variable is given by
\begin{equation}
\frac{dX_k}{dt}=(X_{k+1}-X_{k-2})X_{k-1}-X_k+F
\label{eq:model}
\end{equation}
for $k=1,\ldots,N$. In this equation $F$ represents an external driving, $-X_k$ is a damping 
term, and the quadratic term is an advection term that has been constructed to 
conserve kinetic energy (represented as the sums of squares of $X_k$) in 
the absence of damping. We consider the case of constant external forcing $F$ and where the 
the lattice is periodic in space.  Although other boundary conditions are possible, periodic 
boundaries are of particular relevance for the atmospheric systems of which this model 
was intended to describe.  We will show that the chaotic and periodic solutions we explore 
are composed of traveling structures that travel completely around the periodic-lattice 
many thousands of times. In light of this, other boundary conditions such as an absorbing 
boundary, would have a very strong impact on the dynamics. We have not explored these 
possibilities in detail.

The two parameters $N$ and $F$ completely determine the dynamics. The value of 
$N$ is the system size and $F$ is the magnitude of the external 
forcing. In the following we are interested in the variation of the fractal dimension with these 
parameters. We explore system sizes $N\ge 4$ since $N=3$ does not yield interesting 
dynamics due to the nature of the coupling.  For small values of the forcing $F<8/9$ it has 
been shown~\cite{lorenz:1998} that all solutions decay to the steady solution $X_k=F$. 

We compute the spectrum of Lyapunov exponents $\lambda_k$ and the fractal dimension 
$D_\lambda$ using the standard procedure described in detail in Ref.~\cite{wolf}.  
There are $N$ exponents and for each exponent a set of equations linearized about Eq.~(\ref{eq:model})
are  evolved simultaneously to yield the dynamics of perturbations arbitrarily close to the 
full nonlinear system.  These tangent space equations are
\begin{eqnarray}
	\frac{d\delta X_k^{(i)}}{dt}=X_{k-1}\delta X_{k+1}^{(i)}+(X_{k+1}-X_{k-2})\delta X_{k-1}^{(i)} \nonumber \\
	-X_{k-1}\delta X_{k-2}^{(i)}-\delta X_k^{(i)}
	\label{pert}
\end{eqnarray}
where $\delta X_{k}^{(i)}$ is the $i$th perturbation about lattice site $k$ and $i=1,\ldots,N$. 
The perturbations are reorthonormalized using a Gram-Schmidt procedure after a time $t_N$ to yield the 
magnitude of their growth $\left\| \delta X_k^{(i)}(t_N) \right\|$. Each reorthonormalization 
yields a value of the instantaneous Lyapunov exponent 
\begin{equation}
\tilde{\lambda}_k= \frac{1}{t_N} \ln \left\| \delta X_k^{(i)} \right\|.
\end{equation}
This is repeated and the average value of $\tilde{\lambda}_k$ yields the finite time 
Lyapunov exponent 
\begin{equation}
\lambda_k = \frac{1}{N_t} \sum_{i=1}^{N_t} \tilde{\lambda}_k
\end{equation}
where $N_t$ is the number of reorthonormalizations performed.  The limit $N_t \rightarrow \infty$ yields 
the infinite-time Lyapunov exponent. 

In our numerical simulations, we begin from random initial conditions and use a fourth-order 
Runge-Kutta time integration with a time step $\Delta t=1/64$.  Typically, 
we integrate forward for $1000$ time units before starting the calculation of Lyapunov exponents 
to ensure that all transients have decayed. At this point we begin integrating the 
tangent space equations using $t_N=1$.  All of our reported results are for very long 
simulation times $t \gtrsim 5\times10^5$ and we have computed results for each set of 
parameter values for 10 to 50 different random initial conditions.

\section{Results}

\subsection{The Variation of the Fractal Dimension with System Size}

\subsubsection{Small External Forcing, $F=5$}

\begin{figure}[htb]
\begin{tabular}{cc}
\includegraphics[width=0.25\textwidth]{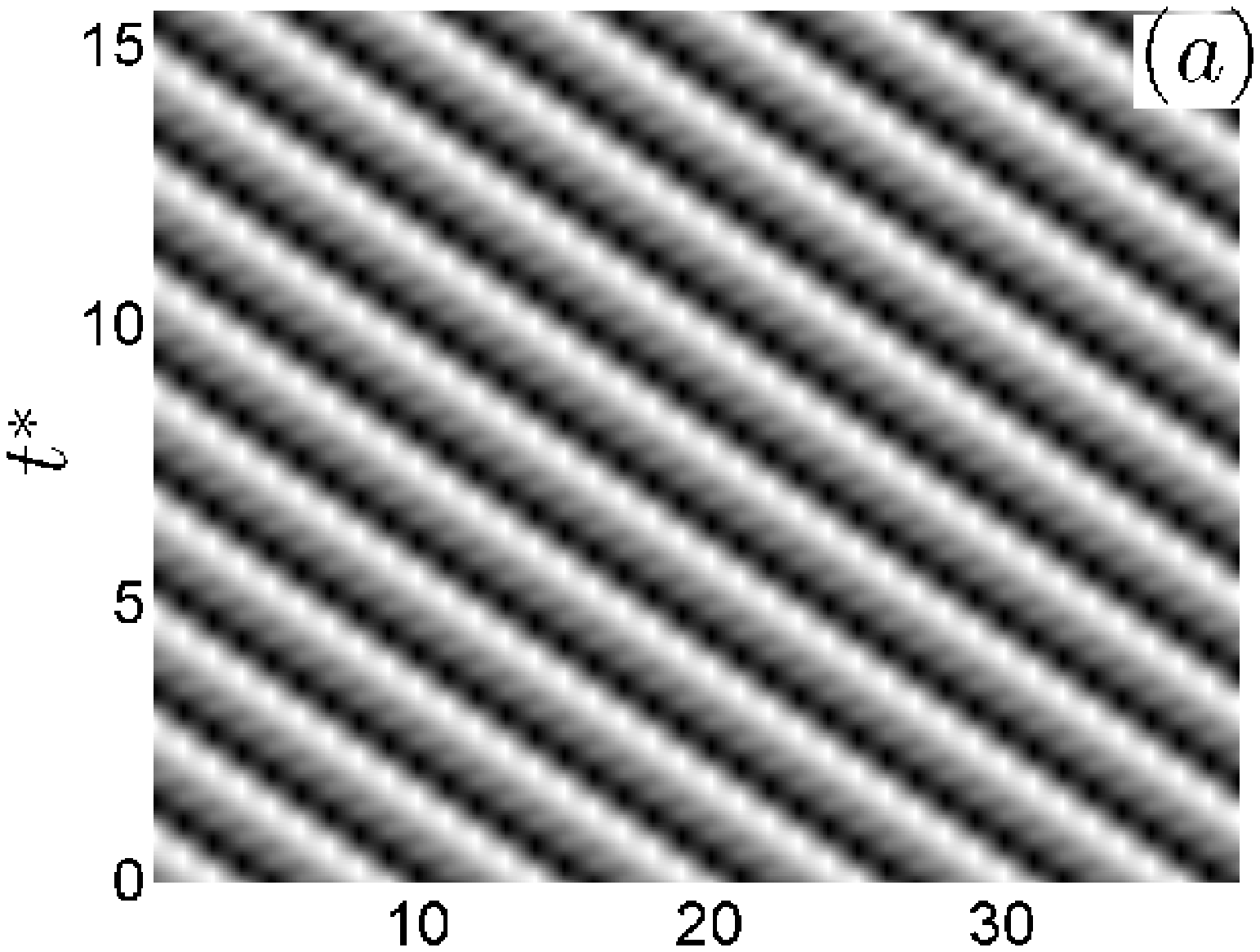} &
\includegraphics[width=0.25\textwidth]{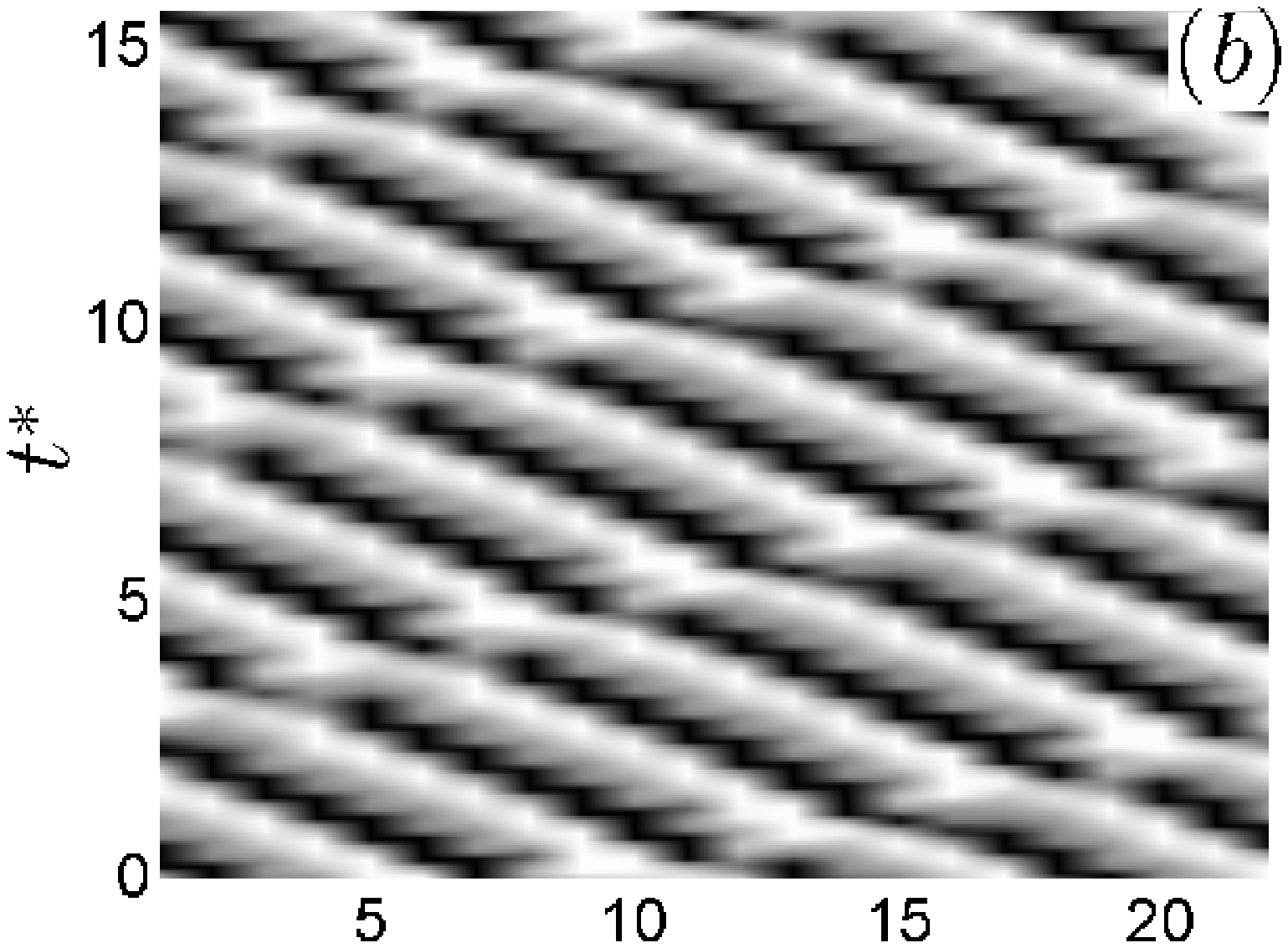} \\
\includegraphics[width=0.25\textwidth]{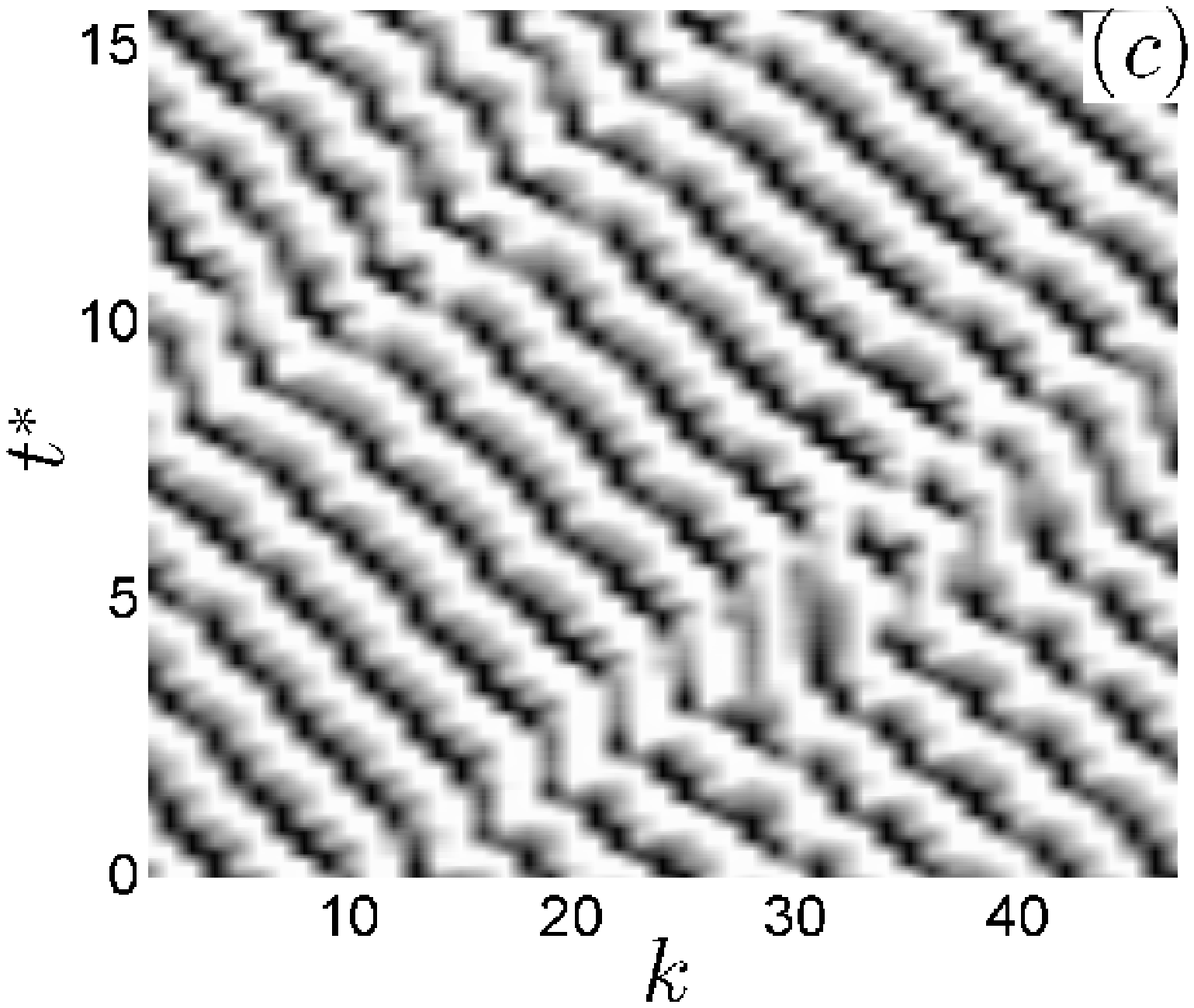} &
\includegraphics[width=0.25\textwidth]{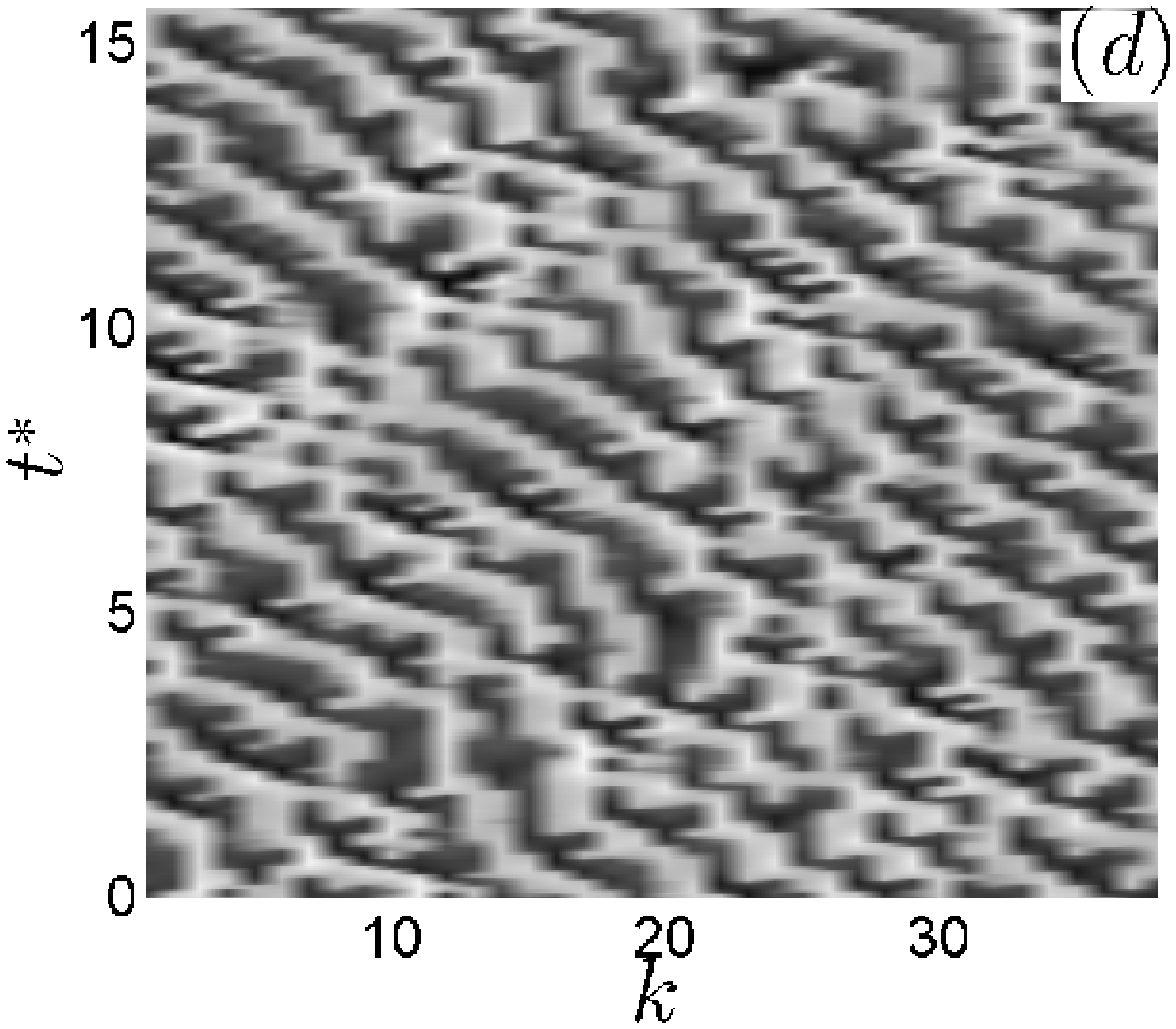}
\end{tabular}
\caption{Space-time plots of $X_k(t)$ for 15 time units after $t = 5.1 \times 10^{5}$ 
indicated by $t^*$. Dark regions are large values and light regions are small values: 
(a)~$N=38$, $F=5$, periodic dynamics; (b)~$N=22$, $F=5$, low-dimensional chaotic 
dynamics; (c)~$N=47$, $F=5$, chaotic dynamics; (d)~$N=38$, $F=10$, chaotic dynamics.}
\label{fig:space-time}
\end{figure}

We first explore the dynamics for a small value of the 
forcing term, $F=5$, over a wide range of system sizes, $4 \le N \le 50$.  We 
find that the dynamics are characterized by windows of periodic and chaotic 
behavior.  Figure~\ref{fig:space-time} shows space-time plots for 
$X_k(t)$ illustrating the variety of dynamics present.   In 
Fig.~\ref{fig:space-time}(a) the dynamics are periodic for $N=38$ yielding 
a wave of constant velocity traveling from right to left. In Fig.~\ref{fig:space-time}(b) 
we show the interesting case $N=22$ where the dynamics are chaotic 
with a small value of the fractal dimension. The dynamics consist of a 
distorted wave structure traveling from right to left.  In Fig.~\ref{fig:space-time}(c) 
chaotic dynamics are shown for $N=47$.  This illustrates the typical chaotic 
dynamics that we have observed where the traveling wave structure  
is still apparent but with significant distortions and 
deviations.

It is evident from these space-time plots that the time for a wave structure to 
travel completely around the periodic lattice ring is $\sim 10$ time units 
for sizes of the lattice rings used here. Our typical simulation time is on the 
order of $10^5$ time units which indicates the number of complete rotations in one of our simulations 
is approximately $10^4$.  The very long simulation times were required to 
gather statistics with sufficient accuracy to address many of the subtle 
questions we study.  Despite the simple nature of the model studied the 
requirement for very long-time simulations is significant. Overall, this 
suggests that the slow and noisy nature of the convergence of the Lyapunov 
spectrum can pose significant computational challenges for more 
complicated models.

\begin{figure}[htb]
\begin{center}
\includegraphics[width=0.45\textwidth]{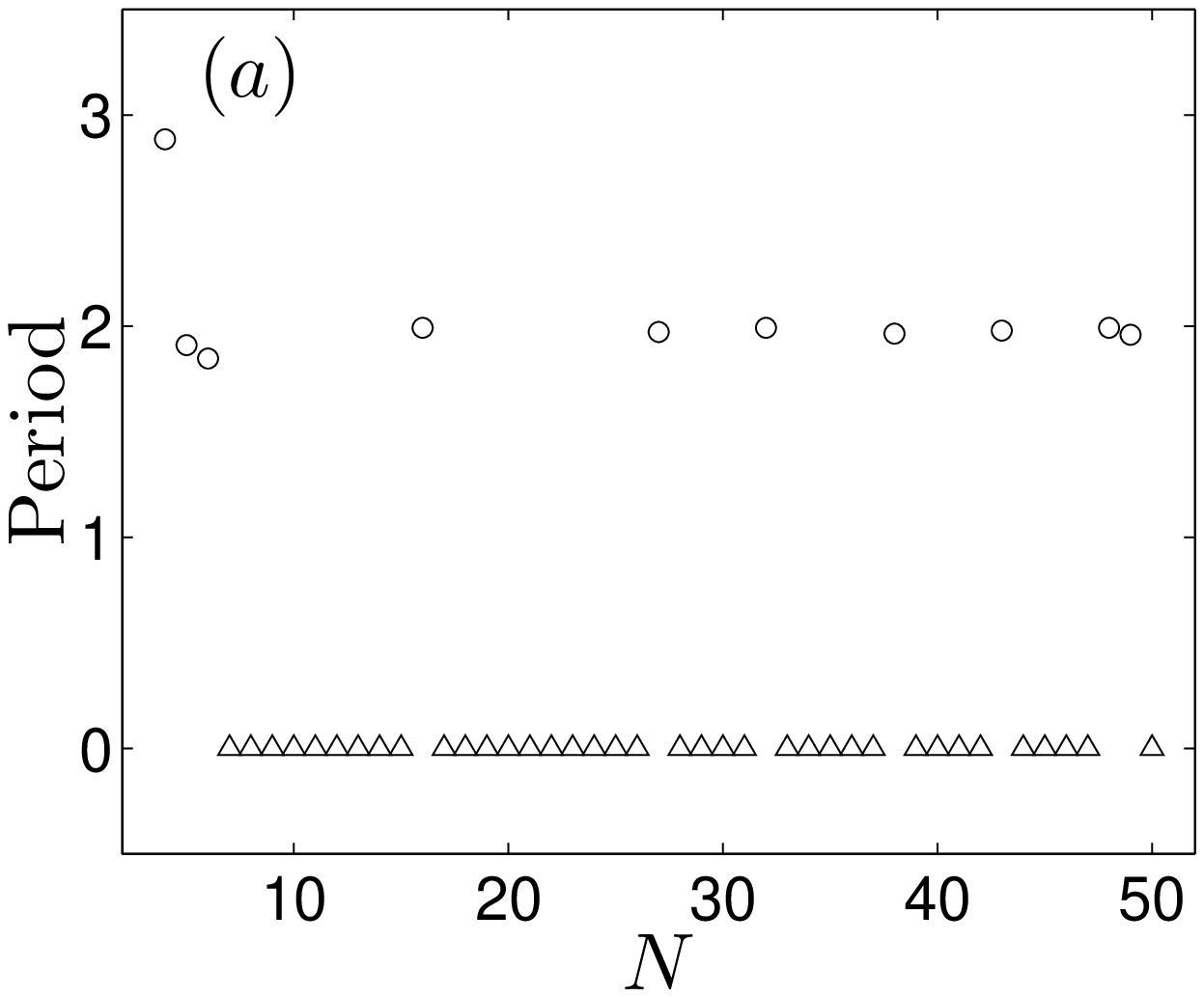}
\includegraphics[width=0.45\textwidth]{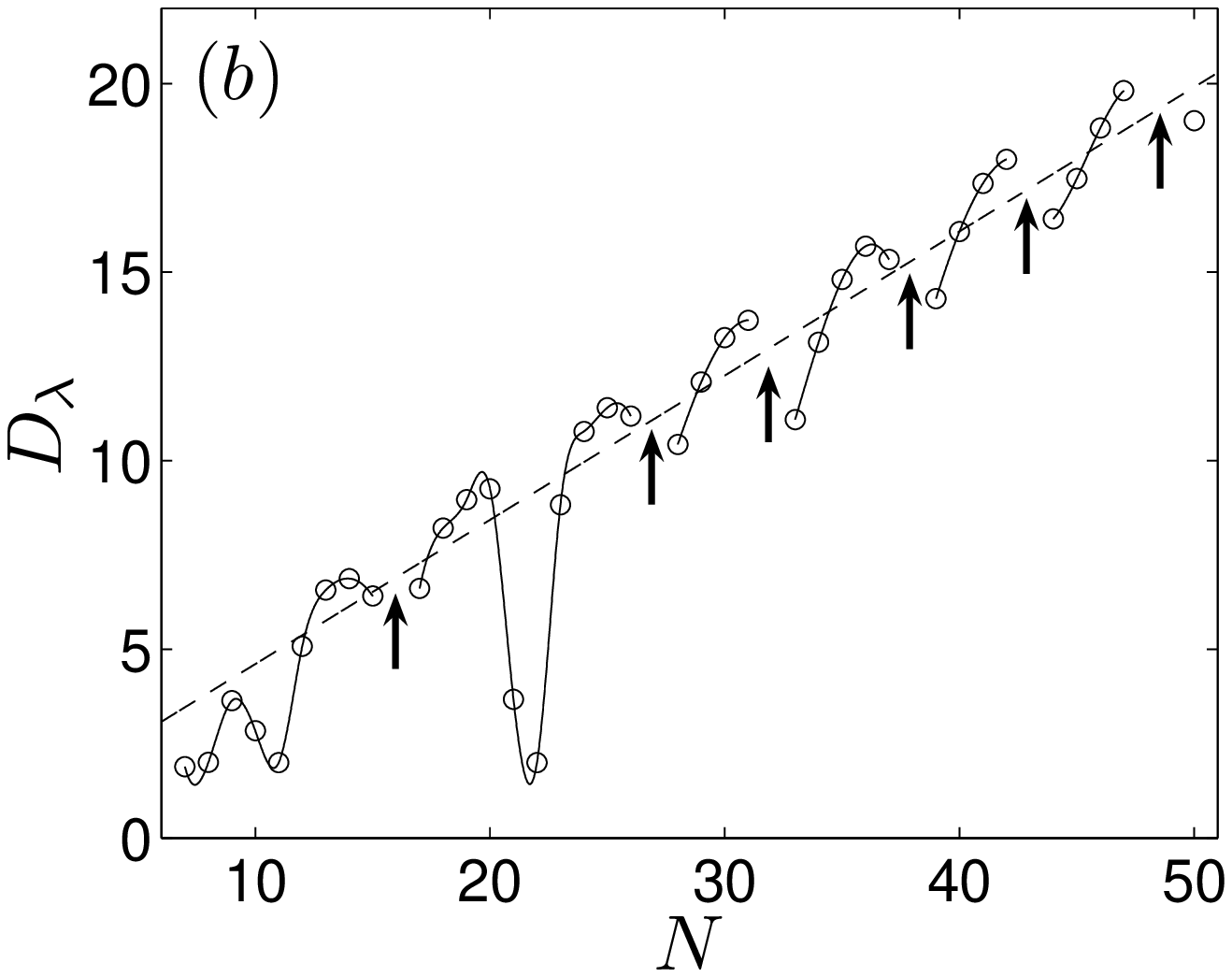}
\end{center}
\caption{The variation of the dynamics with system size $N$ 
for $F=5$. (a)~Periodic dynamics (circles) where the ordinate indicates period 
duration and chaotic dynamics (triangles). Chaotic dynamics have been 
assigned an arbitrary period of `$0$' to fit on a single plot.  (b)~Deviations from 
extensivity illustrated by the variation of $D_\lambda$ with $N$. The circles 
are simulation results and the solid lines 
are to guide the eye.  The maximum value of $\Delta D = 0.77$  
occurs for $N=22$, for $N\ge27$ the maximum deviation is $\Delta D = 0.17$. 
The arrows indicate windows of periodic dynamics.  The error in these calculations 
is $\sim 10^{-2}$ as determined by the standard deviation of $D_\lambda$ using 
$10$ different initial conditions at each $N$. }
\label{fig:F5}
\end{figure}

Figure~\ref{fig:F5}(a) illustrates the variation in dynamics with system size.  The circles 
represent system sizes yielding periodic dynamics and the ordinate is the magnitude 
of the period duration. For systems $N\ge5$ the period of oscillation is approximately 2 
time units. The triangles represent system sizes that yield chaotic dynamics. In order to combine all 
of the results on a single plot the chaotic states were arbitrarily assigned a period of `$0$'.  For 
each system size $N$ we performed 10 long-time numerical simulations starting from different 
random initial conditions.  The type of dynamics found was independent of the initial conditions used. 
The size of the windows of chaotic dynamics are largest for the smaller system sizes. 
For $N\ge27$ the chaotic solutions appear in windows of 4-lattice spacings with the occurrence of a 
single 5-lattice window yielding an average size of $\xi_c = 4.1$. The chaotic dynamics are 
separated by windows of periodicity of 1-lattice spacing with one occurrence 
of a 2-lattice window to give an average size of $\xi_p = 1.2$.
\begin{table}[tbp]
\begin{center}
\begin{tabular}
[c]{l@{\hspace{1.0cm}}l@{\hspace{1.0cm}}l@{\hspace{1.0cm}}l@{\hspace{1.0cm}}l}
                & $\xi_\delta$ & $\xi_0$ & $\xi_L$ \\ \hline \hline
$F = 5$    & 2.6 & 5.7 & 5.2 \\ 
$F = 10$  & 1.35 & 2.0 & 4.5
\end{tabular}
\end{center}
\caption{Characteristic length scales describing the dynamics. The first row are 
results for $F=5$ and the second row are results for $F=10$. $\xi_\delta$ is the chaotic length 
scale found using the fractal dimension,  $\xi_0$ is the average wavelength of the 
deviations in the fractal dimension about extensivity, and $\xi_L$ is the average 
wavelength of the wave structure.}
\label{table:char-lengths}
\end{table}

Figure~\ref{fig:F5}(b) shows the variation of $D_\lambda$ with system 
size $N$.  The circles represent the fractal dimension for system sizes yielding 
chaos and the solid line is to guide the eye.  The error in $D_\lambda$ at each 
value of $N$ is $\sim 10^{-2}$ as determined by the standard deviation in the 
fractal dimension from simulations initiated with 10 different initial conditions. 
The dashed line is a linear curve fit through the data for $N\ge27$ to 
yield an estimate of $D_\text{ext}$ describing 
the dimension of extensive chaos. In all of our calculations of $D_\lambda$ we 
use a 3rd order polynomial curve fit to determine an accurate value for the 
number of Lyapunov exponents that must be included for the sum to vanish 
(the curve fit uses only the 4 sums with the values closest to zero). The 
arrows highlight the gaps between chaotic dynamics indicating 
system sizes yielding periodic dynamics. Using the slope of the dashed line of 
Fig.~\ref{fig:F5}(b) and Eq.~(\ref{eq:chaotic_length}) the chaotic length scale 
is $\xi_\delta = 2.6$.  For reference, the characteristic lengths determined from 
our calculations are collected in Table~\ref{table:char-lengths}. Therefore the 
size of the windows containing chaos are $\xi_c = 1.6 \xi_\delta$ and the size of 
the windows containing periodic dynamics are $\xi_p = 0.46 \xi_\delta$.

To quantify the deviations of the dimension from purely extensive 
chaos we define,
\begin{equation}
 \Delta D=\frac{D-D_{\mathrm{ext}}}{D_{\mathrm{ext}}}.
 \label{eq:deviation}
\end{equation} 	
The overall maximum value of the deviation from extensivity occurs 
for $N=22$ where $\Delta D = 0.77$. If one only considers $N \ge 27$ 
the maximum deviation is $\Delta D = 0.17$.  To estimate the error 
in these calculations we computed results at each value of $N$ using 10 
different random initial conditions.  The standard deviation of the values of $D_\lambda$ 
is $\lesssim 10^{-2}$ which is too small to include as error bars in Fig.~\ref{fig:F5}(b).  
The deviations from extensivity exhibit regular variations about the line of purely extensive 
chaos for $N\ge25$.  To quantify the length scale of these deviations we 
fit a curve through the data symbols to determine the average wavelength of 
these deviations which we will denote as $\xi_0$.  Using this approach yields 
$\xi_0=5.7$.  If each wavelength $\xi_0$ contains a pair of degrees of freedom 
this yields $\xi_0/2=2.85$ lattice spacings for the average volume of a single 
degree of freedom.  It is interesting to point out that $\xi_0/2 \approx 1.1 \xi_{\delta}$ 
suggesting that the deviations from extensivity are due to the addition of new 
chaotic degrees of freedom as the system size is increased. This is similar to 
what has also been observed for the complex Ginzburg-Landau equation~\cite{fishman:2006}. 

\begin{figure}[htb]
\begin{center}
\includegraphics[width=0.45\textwidth]{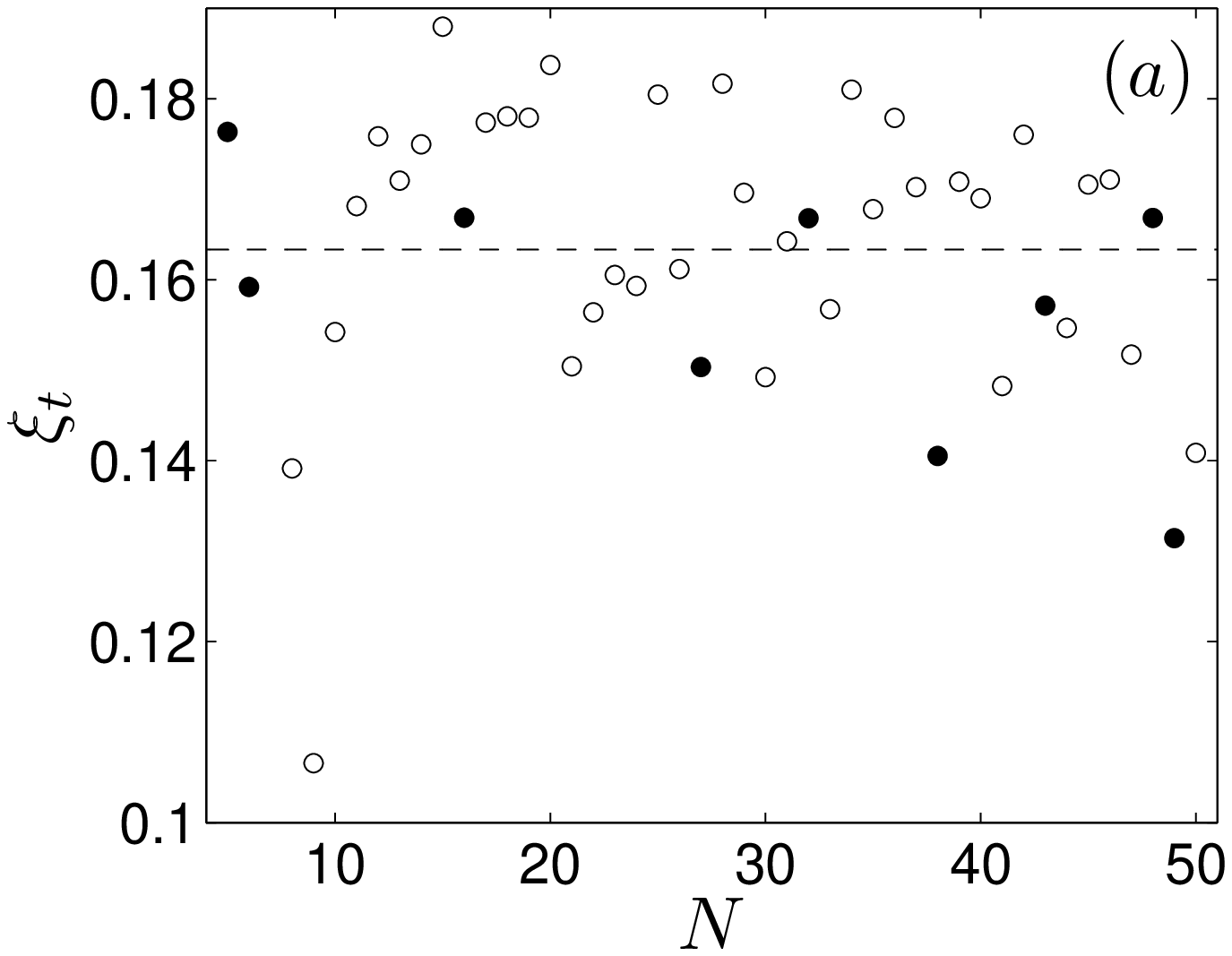}
\includegraphics[width=0.45\textwidth]{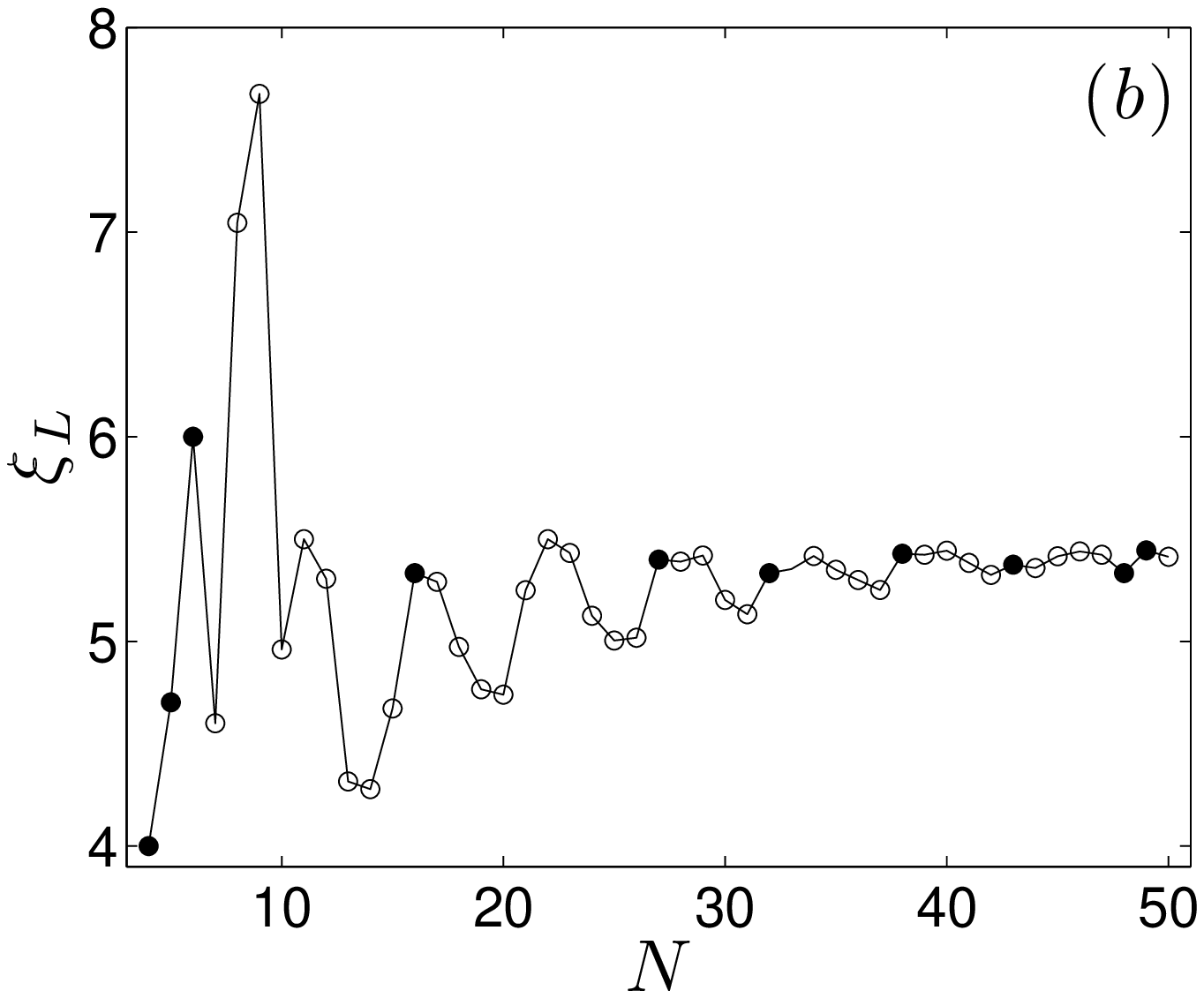}
\end{center}
\caption{The temporal and spatial features of the patterns for $F=5$. (a)~The variation 
of the average correlation time $\xi_t$ with system size $N$. The dashed line is the 
average value of $\xi_t$.  (b)~The variation of the pattern wavelength $\xi_L$ with 
system size. In both panels open symbols represent chaotic dynamics and filled 
symbols represent periodic dynamics.}
\label{fig:patterns_f5}
\end{figure}

We now compare these results based upon the use of Lyapunov exponents with 
what is found by analyzing the spatial and temporal characteristics of the patterns. 
Analysis of the patterns are of particular interest because these measurements 
are often available in experiment whereas the Lyapunov exponent based diagnostics 
are not. Figure~\ref{fig:patterns_f5}(a) shows the 
variation of the correlation time $\xi_t$ with increasing system size.  The correlation 
time is computed from
\begin{equation}
\left< X_k(0) X_k(t)\right> \sim e^{-t/\xi_t}
\end{equation}
where $\left< X_k(0) X_k(t)\right>$ is the autocorrelation of the $k$th variable. The 
results shown are the time average value of the correlation time for the variable $X_1(t)$ 
in the periodic lattice. Each data point is also averaged over simulations begun 
from 10 different random initial conditions.  The open symbols represent chaotic 
dynamics and the filled symbols represent periodic dynamics. The results exhibit 
a scatter about the mean value of $\xi_t = 0.162$ with a coefficient of variation 
of 9.6\% (defined as the standard deviation divided by the mean).  The results 
indicate that the time dynamics remain relatively constant as the system size 
increases.

To quantify the spatial variation of the pattern we have computed the average 
pattern wavelength $\xi_L$ with increasing system size as illustrated in 
Figure~\ref{fig:patterns_f5}(b).  The wavelength is computed in Fourier space 
using the location of the largest peak at small wavenumber to capture the 
size of the basic wave structure and is averaged over all time for each 
initial condition. Computation of the two-point spatial correlation length is 
difficult due the lack of an exponential decay in the correlation functions for the 
parameters we have explored. For smaller systems $N \lesssim 10$ the deviations are 
largest whereas for $N>10$ the wavelength varies about an average value of 
$\xi_\lambda \approx 5.2$ lattice spacings.  For the chaotic solutions, the fluctuations 
about the average value over the 10 different initial conditions yields a coefficient 
of variation of approximately $14\%$. For the periodic dynamics the wavelength of the periodic 
states were nearly identical to the precision of our calculations over the different 
initial conditions.

Our space-time diagnostics indicate that with increasing system size the dynamics 
tend toward a state where $\xi_t \approx 0.16$ and $\xi_L \approx 5.2$. The 
wavelength of the pattern in the large system limit is $\xi_L \approx 2 \xi_\delta \approx 2.2 \xi_0$ 
indicating that each wavelength of the pattern contains approximately 2 chaotic 
degrees of freedom on average. In this case, it is useful to compare the deviations 
from extensivity with the variations in the wavelength of the patterns for increasing 
system size.

\begin{figure}[htb]
\begin{center}
\includegraphics[width=0.5\textwidth]{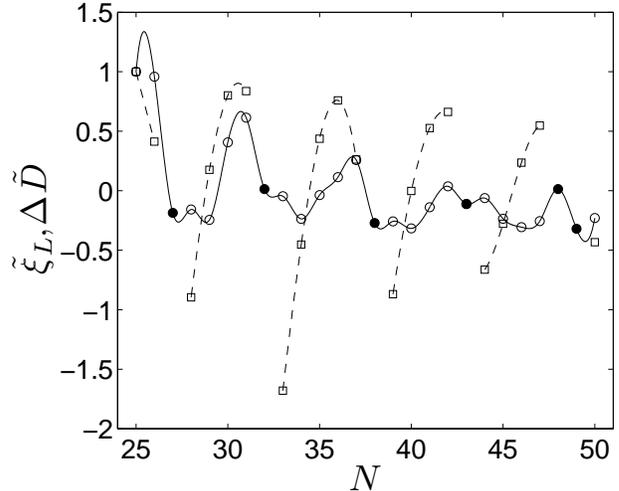}
\end{center}
\caption{A comparison of the variation in the normalized pattern wavelength $\tilde{\xi}_L$ 
and the normalized deviation from extensivity $\Delta \tilde{D}$ for $F=5$ and $N \ge 25$. 
The constant normalization factor is chosen such that $\Delta \tilde{D}$ and $\tilde{\xi}_L$ 
each equal unity for $N=25$ allowing both curves to fit on a single plot.  The wavelength 
results are given by the circles and the solid line (open symbols are chaotic dynamics, 
filled symbols are periodic dynamics). The deviations from extensivity are 
shown by the square symbols and the dashed line where gaps indicate regions 
of periodic solutions.  In both cases the lines are curve fits to guide the eye.}
\label{fig:wavelength_dimension}
\end{figure}

The variation of $\xi_L$ about its mean value is quite similar to what is found for the 
deviations from extensivity $\Delta D$ as illustrated in Fig.~\ref{fig:wavelength_dimension}. 
We plot the variation of the normalized wavelength $\tilde{\xi}_L$ and the normalized 
deviations from extensivity $\Delta \tilde{D}$ with the system size.  The constant 
normalization factors are chosen such that the normalized wavelength and deviation 
from extensivity equal unity for $N=25$ allowing both curves to be shown on 
a single plot.  The circles are results for the wavelength of the pattern where 
solid symbols indicate periodic dynamics, open symbols represent chaotic 
dynamics, and the solid line is a curve fit.  The dashed line and 
the square symbols represent the deviations from extensivity.  Gaps in the 
dashed line occur for system sizes exhibiting periodic dynamics.

Figure~\ref{fig:wavelength_dimension} illustrates a correlation between 
the pattern wavelength and the deviation from extensivity. The 
general trend is that both the fractal dimension and the pattern wavelength increase  
with increasing system size until the pattern adjusts by adding an additional wave structure  
to the periodic lattice effectively reducing the average wavelength of the pattern. 
With the addition of the wave structure the dynamics are periodic and the fractal dimension vanishes.  A similar 
trend is seen in the work of Fishman {\em et al.}~\cite{fishman:2006} where periodic dynamics were 
found for system sizes where the deviation from extenisivity was predicted to be a 
minimum for the one-dimensional complex Ginzburg-Landau equation.

\subsubsection{Intermediate External Forcing, $F=10$}

We next explore the dynamics for a larger value of the forcing $F=10$ over the  
range of system sizes $5 \le N \le 50$.  For each system size we have performed 
50 independent numerical simulations starting from different random initial conditions. 
Each simulation was allowed to continue until $t = 5 \times 10^{5}$ and in computing 
our results we only used data in in the time interval $3 \times 10^{5} \le t \le 5 \times 10^{5}$ to ensure 
the decay of all transients and in order to gather good statistics.  We found chaotic 
dynamics at every system size and for each random initial condition. A representative 
space-time plot for our results is shown in Fig.~\ref{fig:space-time}(d) for the 
case of $N=38$.  The patterns consist of distorted wave structures traveling from 
right to left.

\begin{figure}[tb]
  \begin{center}
    \includegraphics[width=0.45\textwidth]{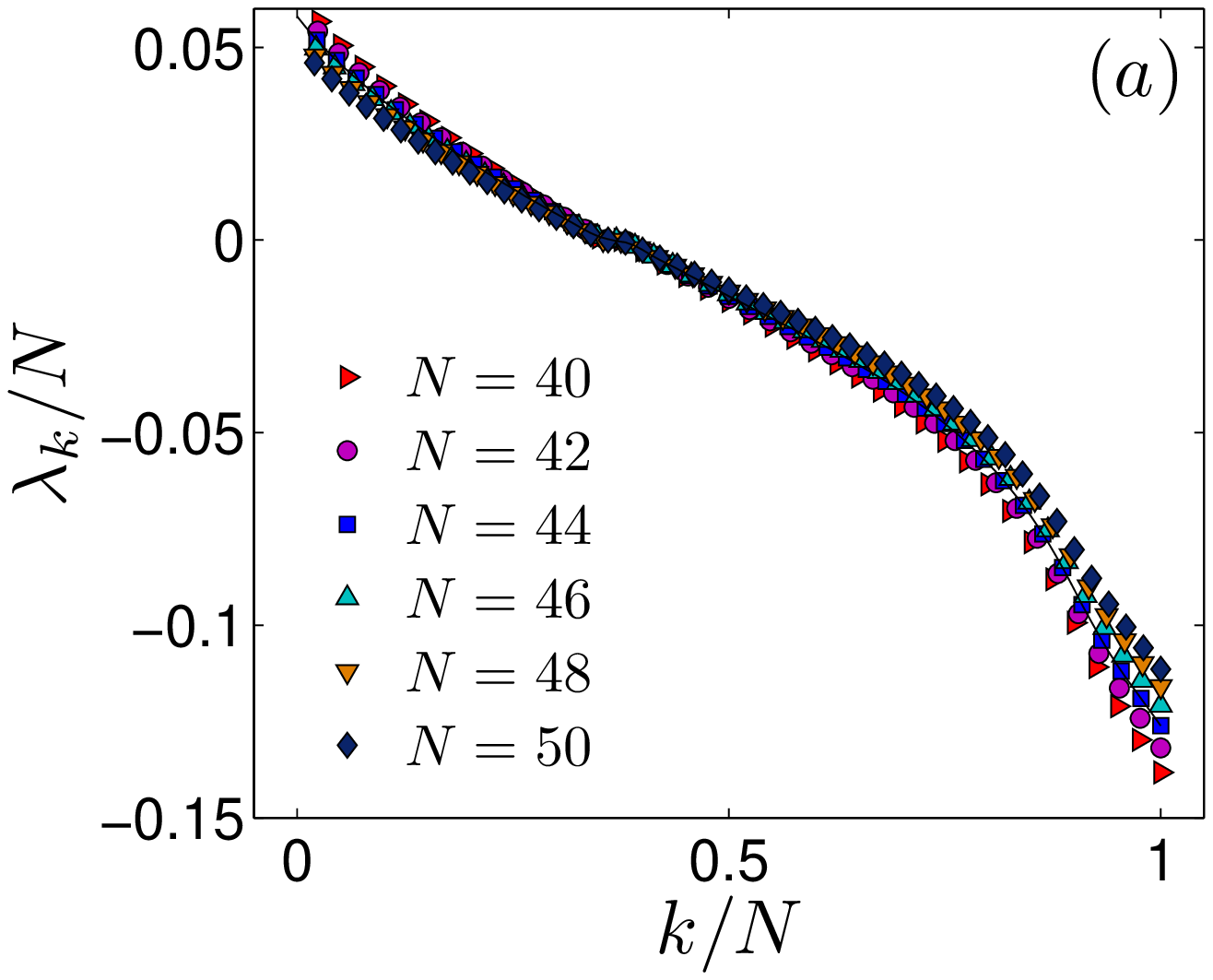}
    \includegraphics[width=0.45\textwidth]{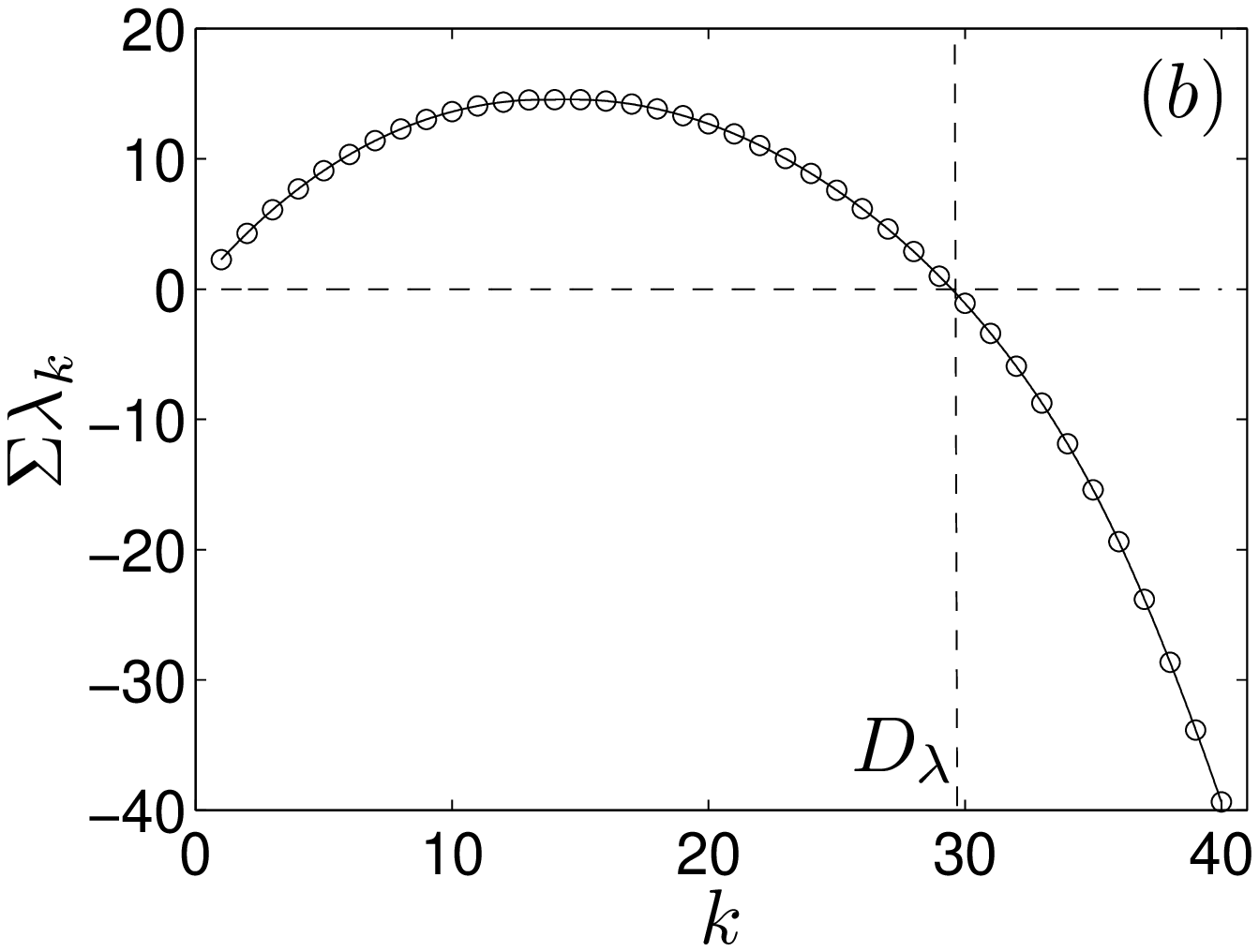}
  \end{center}
  \caption{(color online) (a)~The variation of the Lyapunov exponents $\lambda_k/N$ 
  with $k/N$ for $k=1,\ldots,N$ where $N$ is the system size and $F=10$. 
  The solid line is the average value using all of the data. (b)~The variation 
  of the summation of the exponents with $k$ for $F=10$ and $N=40$.  The fractal 
  dimension $D_\lambda$ is the value where the summation vanishes. The 
  solid line is a 6th order polynomial curve fit and the fractal dimension $D_\lambda$ 
  is shown by the vertical dashed line.}
  \label{fig:spec}
\end{figure}

The spectra of Lyapunov exponents are shown in Fig.~\ref{fig:spec}(a) for six 
different values of system size. The results are normalized by $N$ and for 
extensive chaos the spectra collapse onto a single curve as expected. The 
solid line is the average of the results for the different system sizes shown. 
The variation of the summation of the exponents with the number of exponents is 
illustrated in Fig.~\ref{fig:spec}(b) for the case of $N=40$.  The solid line is a 6th order 
polynomial curve fit through the average values of the data.  The fractal dimension 
is indicated by the vertical dashed line at the location where the summation 
vanishes.

\begin{figure}[tb]
  \begin{center}
    \includegraphics[width=0.45\textwidth]{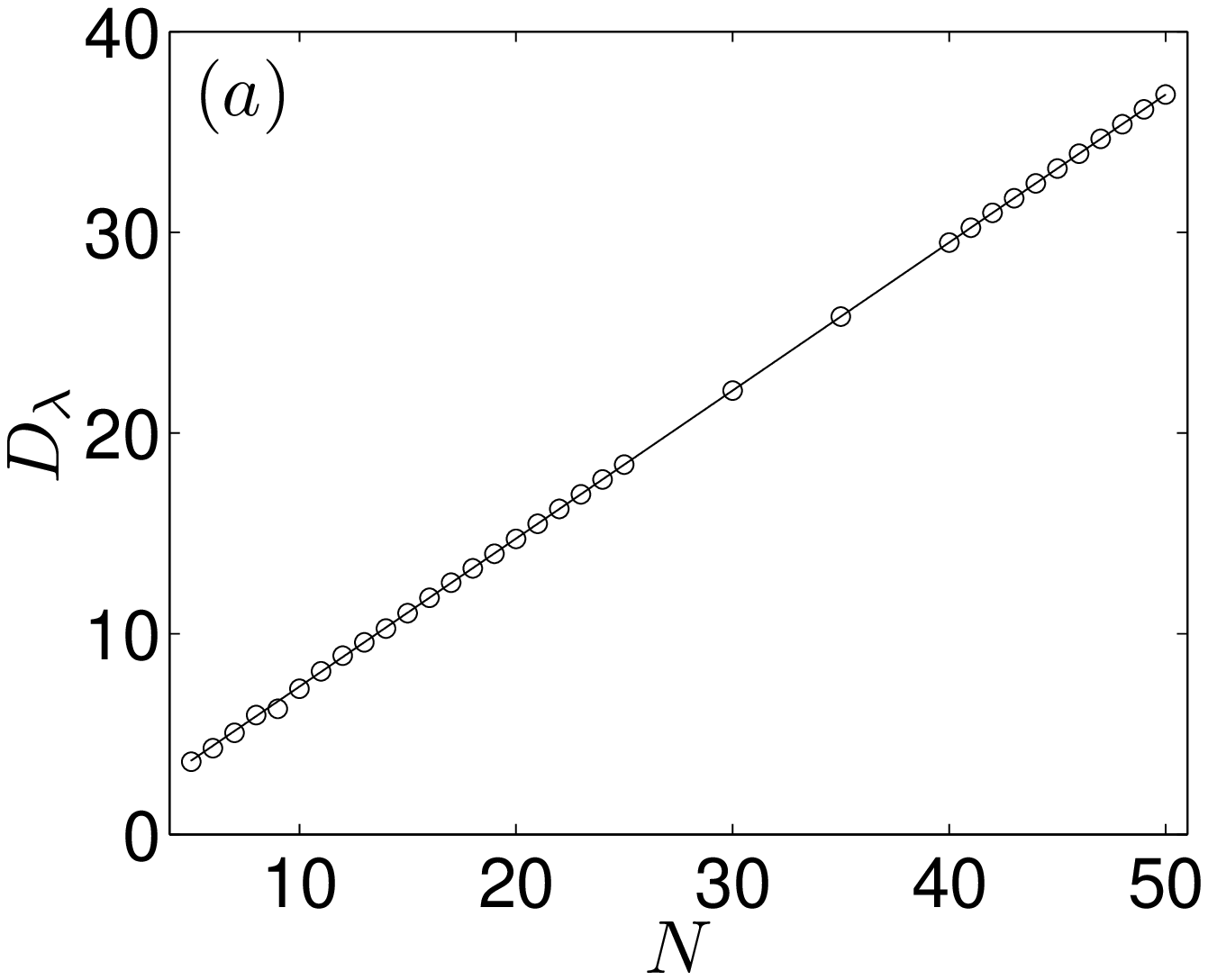}
    \includegraphics[width=0.45\textwidth]{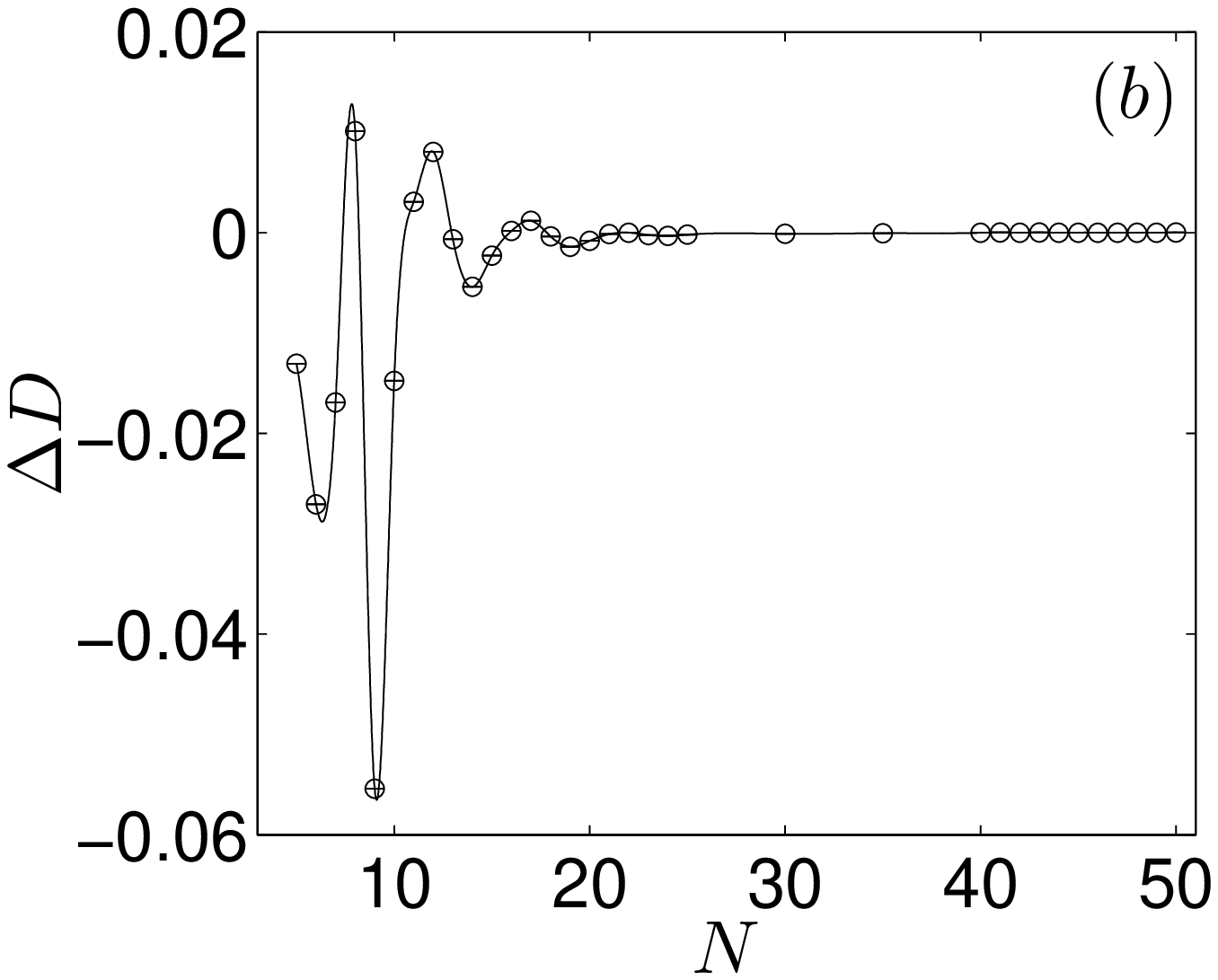}
  \end{center}
  \caption{The variation of the dynamics with the system size for $F=10$.  (a)~The 
variation of $D_\lambda$ with $N$, the solid line is a linear curve fit through the data 
indicating extensive chaos. For $N=4$ the dynamics are periodic 
and are not included in the curve fit. (b)~The variation of the deviation from 
extensivity $\Delta D$ with $N$. The circles are results from numerical simulation and the 
solid line is a curve fit to guide the eye. The deviations from extensivity are on the order of 
5\% and decreases with increasing $N$.  The fluctuations about the average values 
shown in each panel, as determined from the standard deviation of results from 50 different 
initial conditions, is on the order of $\sim 10^{-5}$.}
\label{fig:deviations}
\end{figure}

The variation of the fractal dimension with system size is shown in Fig.~\ref{fig:deviations}. 
Figure~\ref{fig:deviations}(a) illustrates the variation of $D_\lambda$ with $N$ where the 
symbols represent the average value over the 50 different initial conditions.  An estimate 
of the error in these calculations is the standard deviation of the values of the dimension 
at each system size.  Using this, the error is found to be quite small with a magnitude 
of $\sim 10^{-5}$ over the entire range of system sizes. The solid line is a linear curve fit 
through the data points indicating extensivitiy. This yields a value of $D_{\text{ext}}$ for  
any system size $N$ and using Eq.~(\ref{eq:chaotic_length}) yields a  value of the natural 
chaotic length scale of $\xi_\delta=1.35$.  It is important to point out that since $\xi_\delta > 1$, 
incremental changes in $N$ by adding a single lattice site allow the variation in the fractal 
dimension to be observed for changes in system size that are smaller than the chaotic 
length scale. 

The deviations of the fractal dimension about extensivity $\Delta D$ are shown in 
Fig.~\ref{fig:deviations}(b) where we have used Eq.~(\ref{eq:deviation}) and the solid line 
is to guide the eye.  The magnitude of the deviation from extensivity for $N \le 25$ is 
$\sim5\%$ at its largest value and tends towards zero with increasing $N$. Error bars are included 
and, over the entire range of system sizes, have a magnitude on the order of $\sim 10^{-5}$ 
which is 3 orders of magnitude smaller than the absolute values of $\Delta D$.  Due to 
the very small magnitude of our error bars we can discern these deviations in the 
fractal dimension from extensivity for all values of the system size. This suggests that 
the deviations from extensivity are not entirely finite-size effects and are inherent to 
the underlying chaotic dynamics. For the larger system sizes $40 \le N \le 50$ the 
magnitude of the deviations from extensivity reduce to $\Delta D \approx 10^{-5}$. 
We note that over this range the magnitude of the error is on the order of $\sim 10^{-6}$ and 
these small deviations are still an order of magnitude larger than the error bars.

Figure~\ref{fig:deviations}(b) illustrates that the variation of $\Delta D$ with $N$ is 
non-monotonic in its oscillating decay in amplitude toward extensivity.  The previous 
studies exhibiting deviations from extensivity of O'Hern~\emph{et al.}~\cite{ohern:1996} 
and Fishman~\emph{et al.}~\cite{fishman:2006} both observed a rapid monotonic decay 
towards extensive chaos. At present we do not have a good understanding of the 
physical origin of the non-monotonicity in our results which is perhaps related to the 
discrete spatial structure of the Lorenz-96 model.

It is useful to separate these findings into two regimes, a small system regime 
where $N \le 25$ and a large system regime where $N\ge40$. The small system 
regime includes the onset of extensivity whereas the large systems are extensive. 
In a more complicated system, one would typically only have access to the deviations 
from extensivity for the small systems sizes since the deviations are largest in this 
range.  However, this is also the regime where one could expect finite size 
effects to also be important and complicate the results.  The large system limit 
is expected to be less influenced by these finite size effects and to provide a better 
estimate of the length scale of a chaotic degree of freedom.

Over these two regimes we find that the wavelength of the deviations varies 
significantly.   To quantify this we compute the zero crossings of $\Delta D$ over 
each range to estimate the wavelength of oscillation about $D_{\text{ext}}$ given 
by $\xi_0$.  For $N \le 25$ there are 3 wavelengths with values of $(3.3, 5.3,1.9)$ 
that yield an average of $\xi_0 \approx 3.5$. Assuming each wavelength of $\Delta D$ 
corresponds to the addition of two degrees of freedom yields $\xi_0/2 \approx 1.75$ 
for the size of a single degree of freedom. Comparison of this length with the natural 
chaotic length scale yields that the size of a single degree of freedom is approximately 
$1.3 \xi_\delta$.  The large spread in the values of the wavelengths suggests 
some competition with pattern selection mechanisms.

In the large system limit $N \ge 40$ there are 4 wavelengths characterizing the deviations 
from extensivity with values $(1.7,1.7,2.6,2.3)$ yielding an average value of 
$\xi_0 \approx 2$.  The magnitude and variation of the wavelengths are smaller than 
what was found for the smaller systems.  Using the same arguments as before yields 
that the size of a single degree of freedom is  $\xi_0/2 \approx 1$. Comparison 
with the natural chaotic length scale yields that a single degree of freedom 
is $0.74 \xi_\delta$.  As in the case of $F=5$ the results yield that $\xi_0/2 \approx \xi_\delta$ 
suggesting that the natural chaotic length scale and the wavelength of the deviations 
from extensivity are related.
\begin{figure}[htb]
\begin{center}
\includegraphics[width=0.45\textwidth]{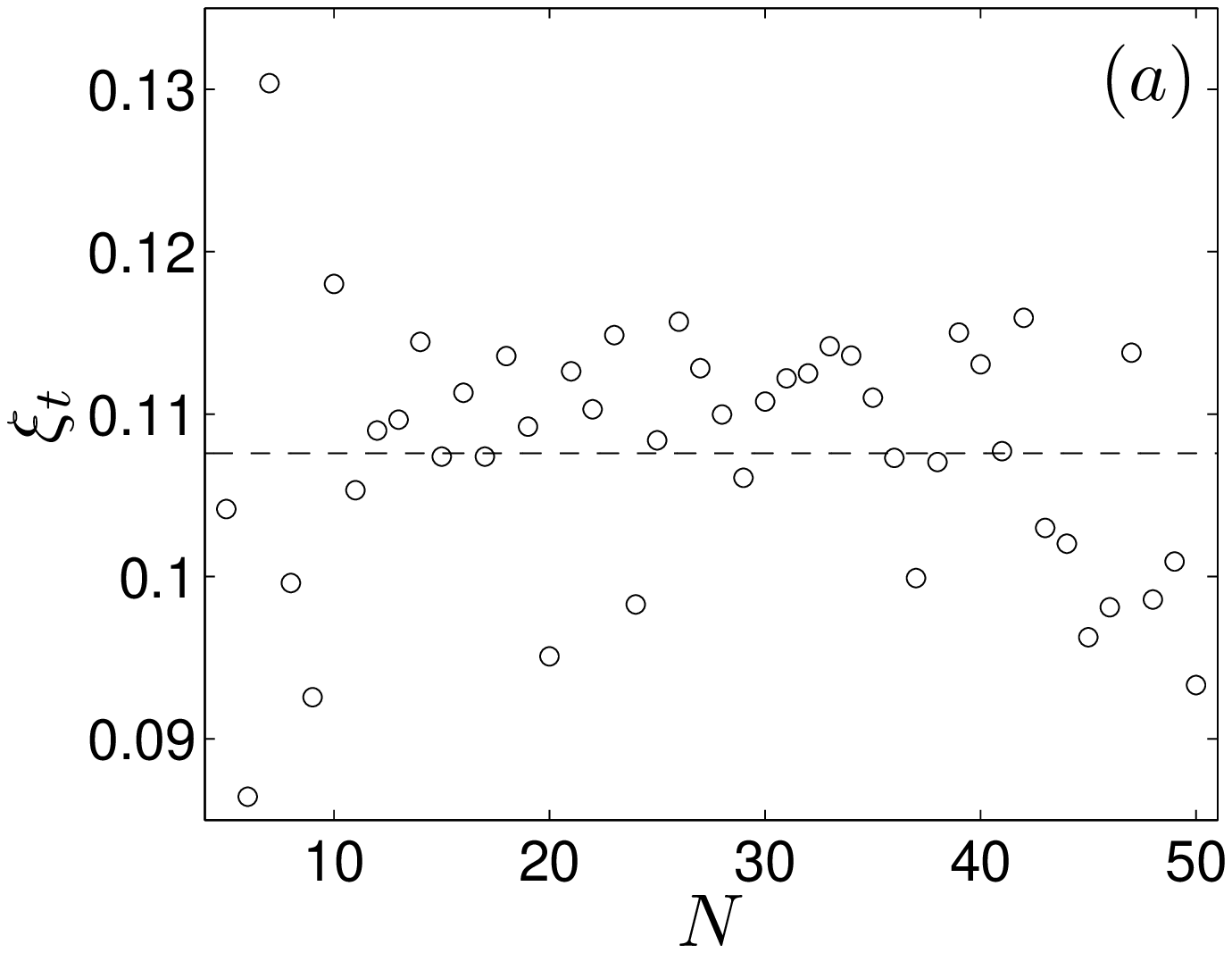}
\includegraphics[width=0.45\textwidth]{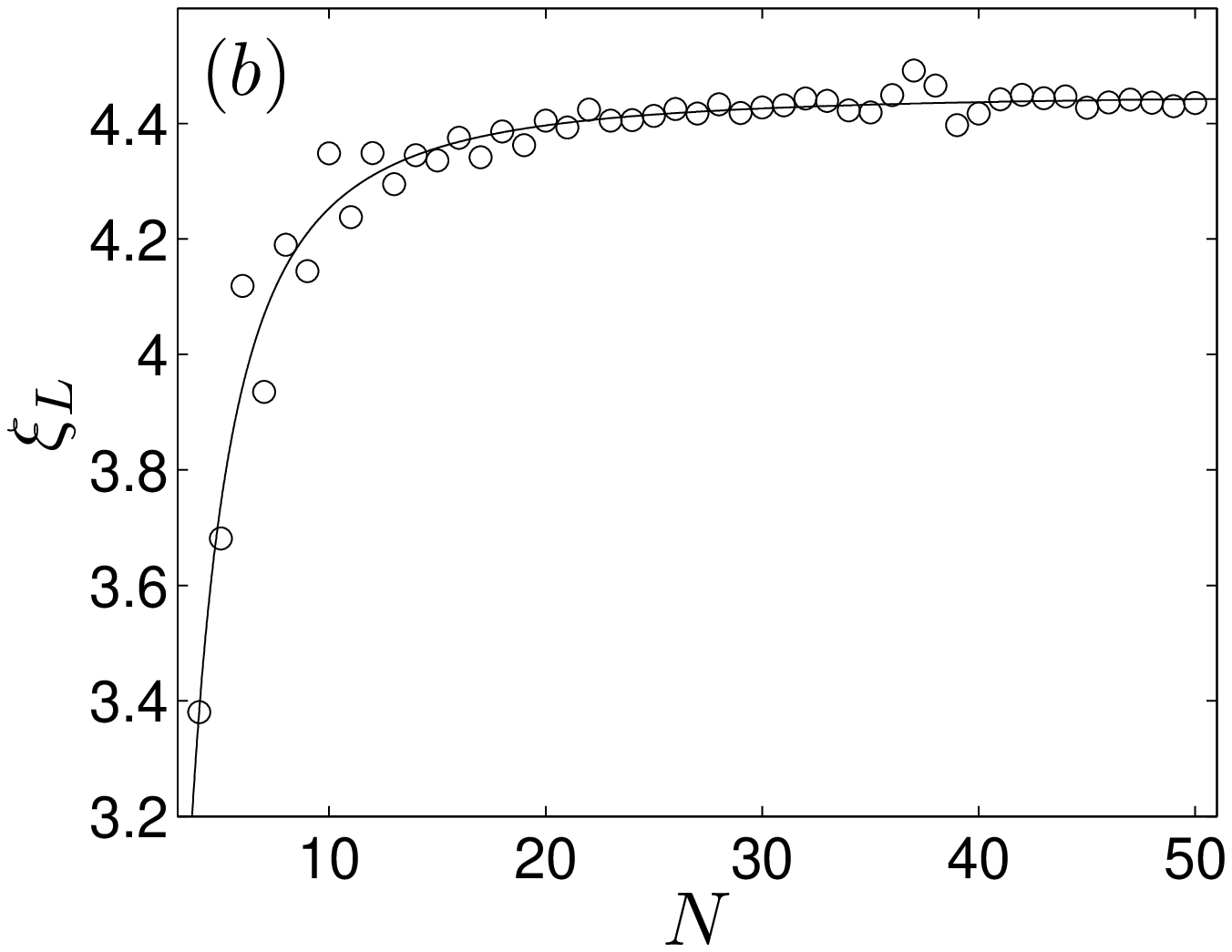}
\end{center}
\caption{The temporal and spatial characteristics of the patterns for $F=10$. (a)~The 
variation of the average correlation time $\xi_t$ with system size $N$. The dashed line 
is the average value of $\xi_t$. (b)~The variation of the average pattern wavelength 
$\xi_L$ with system size.  The solid line is a curve fit of the form 
$\xi_L = 4.45 - 13.6 N^{-1.83}$.}
\label{fig:patterns_f10}
\end{figure}

In order to better characterize the patterns we have also computed characteristic time and 
length scales describing their dynamics.  Our motivation is to provide 
further insight into the transition between the small and large system limits in order 
to separate finite size effects from dynamical effects related to extensivity.  We are 
also interested in quantifying any relationship between the Lyapunov based 
diagnostics and the pattern dynamics. Figure~\ref{fig:patterns_f10}(a) shows 
the variation of the average correlation time $\xi_t$ with $N$ over 50 different 
initial conditions.  The time dynamics are noisy and the correlation time has an 
average value of approximately $\xi_t \approx 0.11$ (indicated by the dashed line).  
The time dynamics are faster than what was found for $F=5$ as expected for the 
increased value of the forcing term.

Figure~\ref{fig:patterns_f10}(b) illustrates the variation in the average wavelength 
of the patterns with $N$ over the different initial condtions. The wavelength 
variation shows two regions of interest. For small system sizes $N \lesssim 20$ 
the pattern wavelength increases rapidly. For larger systems the wavelength 
remains relatively constant with an average value of $\xi_\lambda \approx 4.5$.  
The region of increasing wavelength corresponds to 
system sizes where the pattern selection is affected significantly by the size 
of the domain. This corresponds roughly with the region where the fractal 
dimension is approaching extensivity in Fig.~\ref{fig:deviations}(b).  These trends 
are well captured by a power-law, the solid line in the figure is a curve fit through 
the data of the form $\xi_L = 4.45 - 13.6 N^{-1.83}$.  The fluctuations of the 
wavelength about the mean value for these chaotic states yields a coefficient of 
variation of approximately $30\%$.

These results suggest that our measurements of the deviations in the 
fractal dimension for the large system limit are quantifying the chaotic 
dynamics in a regime that is not strongly affected by finite size effects.  Our 
space-time diagnostics indicate that the pattern dynamics, in the large system 
limit, have temporal variations with $\xi_t = 0.106$ and a spatial wavelength 
on the order of $\xi_L = 4.45$.  In this case, $\xi_L = 3.3 \xi_\delta$ indicating 
that a wavelength of the pattern contains, on average, 3.3 chaotic degrees 
of freedom.  A more detailed exploration of the relationship between $\xi_L$ and $\xi_\delta$ 
for increasing values of the forcing $F$ will be explored in the following section.  

\subsection{Variation of the Fractal Dimension with Forcing}

In the previous sections we have considered the large system limit, defined as the limit 
where the chaotic degrees of freedom are much smaller than the system size.  This 
was achieved by holding the external forcing $F$ fixed while the system size $N$ was
increased.  This is the typical manner to study spatiotemporal chaos and 
has been referred to as the `spatiotemporal chaos' limit~\cite{cross:1993}. For 
example, in a Rayleigh-B\'{e}nard 
convection experiment this could be accomplished by holding the Rayleigh number 
constant while increasing the aspect ratio of the convective domain.

However, it is also possible to explore the large system limit by keeping the system 
size fixed while increasing the external forcing. This has been referred to as the strong driving 
or `strong turbulence' limit~\cite{cross:1993}. In this case, it is expected that the chaotic 
degrees of freedom will become smaller with increasing forcing to yield the large system 
limit.  In a Rayleigh-B\'{e}nard convection experiment 
this would be accomplished by increasing the Rayleigh number in a convection 
domain of fixed size.  It has been conjectured that the fractal dimension will also 
exhibit a power-law dependence with respect to the value of the forcing~\cite{cross:1993}. 

We have explored this strong driving limit in the Lorenz-96 model by performing a series 
of simulations for increasing values of $F$ while the system size $N$ is held constant. We 
studied 6 values of the forcing over the range $5 \le F \le 30$. For any particular value of 
$F$ we performed 10 numerical simulations starting from different random initial conditions 
and allowed the simulation to run for $5 \times 10^5$ time units to ensure good statistics.  
We computed these results for 5 different system sizes where $15 \le N \le 35$.

\begin{figure}[htb]
  \begin{center}
    \includegraphics[width=0.5\textwidth]{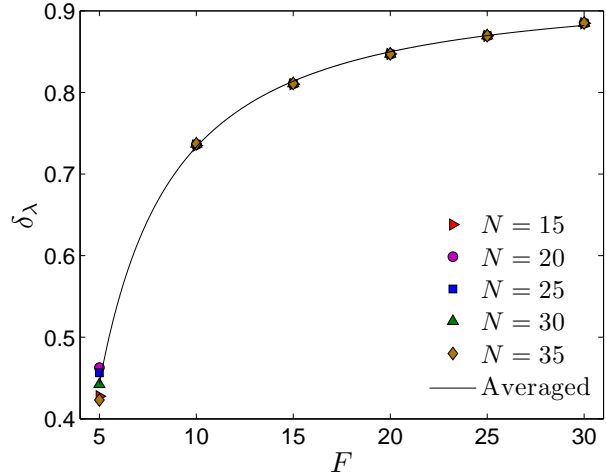}
  \end{center}
\caption{(color online) The variation of the dimension density $\delta_\lambda$ with 
the forcing $F$ and system size $N$. The results are for 5 different system sizes indicated 
in the legend. The symbols are simulation results and the solid line is a power-law curve 
fit through the average value of the data. Each symbol is the average value from results 
computed using 10 different initial conditions. The standard deviation of these results for 
each symbol is $\sim 10^{-5}$.}
\label{fig:DF}
\end{figure}

It is more convenient to discuss these results using the intensive dimension density,
\begin{equation}
\delta_\lambda = \frac{D_\lambda}{N}.
\end{equation}
The variation of the dimension density with external forcing is shown in Fig.~\ref{fig:DF}. 
Each symbol is the average value of the 10 simulations from different random 
initial conditions.  The standard deviation of the results are $\sim 10^{-5}$ and 
have not been included as error bars due to their small magnitude.

The dimension density exhibits power-law behavior and the results for different values of 
$N$ collapse onto a single power-law curve given by, 
\begin{equation}
\delta_\lambda(F) = a + b F^{-\alpha}
\label{eq:powerlaw}
\end{equation}
where $a = 0.93$, $b=-4.03$, and $\alpha=1.32$.  The power-law curve fit is 
given by the solid line. The coefficient $a$ in this 
case is the predicted value of the dimension density for external forcing of 
infinite magnitude $\delta_\infty$. The scatter in the results is quite small 
for all values of $F$ explored with the largest deviations occurring at the 
smallest value of the forcing.  The coefficients of the power-law are 
also collected in Table~\ref{table:powerlaws} for reference.

The inverse of the dimension density is simply the natural chaotic length scale 
$\xi_\delta = \delta_\lambda^{-1}$. Therefore Fig.~\ref{fig:DF} can be used to 
illustrate the manner in which the chaotic length scale decreases with increasing 
forcing. This is a spatially discrete system with $N$ degrees of freedom and 
the theoretical limit for the dimension density is $\delta_\lambda=1$ corresponding to 
a chaotic length scale of a single lattice spacing $\xi_\delta=1$.  Using 
our results for $\delta_\infty$ yields a value of $\xi_\delta=1.08$ for the chaotic length in the limit 
of infinite forcing. This is in contrast to a fluid system described by partial differential 
equations with an infinite number of degrees of freedom.  It is interesting to note that 
the variation of the fractal dimension with the degree of external forcing has been 
explored numerically for turbulent Rayleigh-B\'{e}nard convection to yield a linear 
dependence~\cite{sirovich:1991}.

\begin{figure}[htb]
  \begin{center}
    \includegraphics[width=0.45\textwidth]{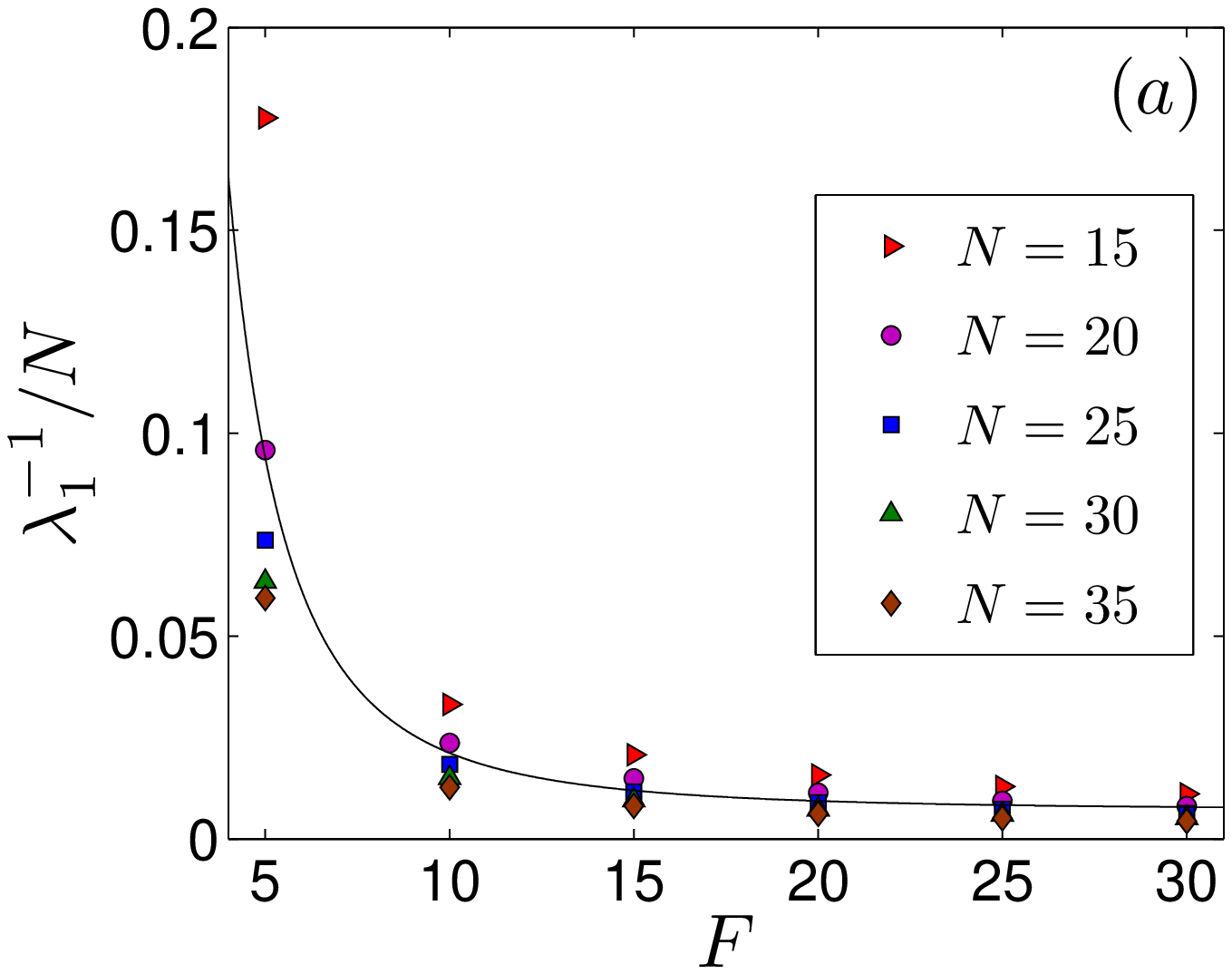}
    \includegraphics[width=0.45\textwidth]{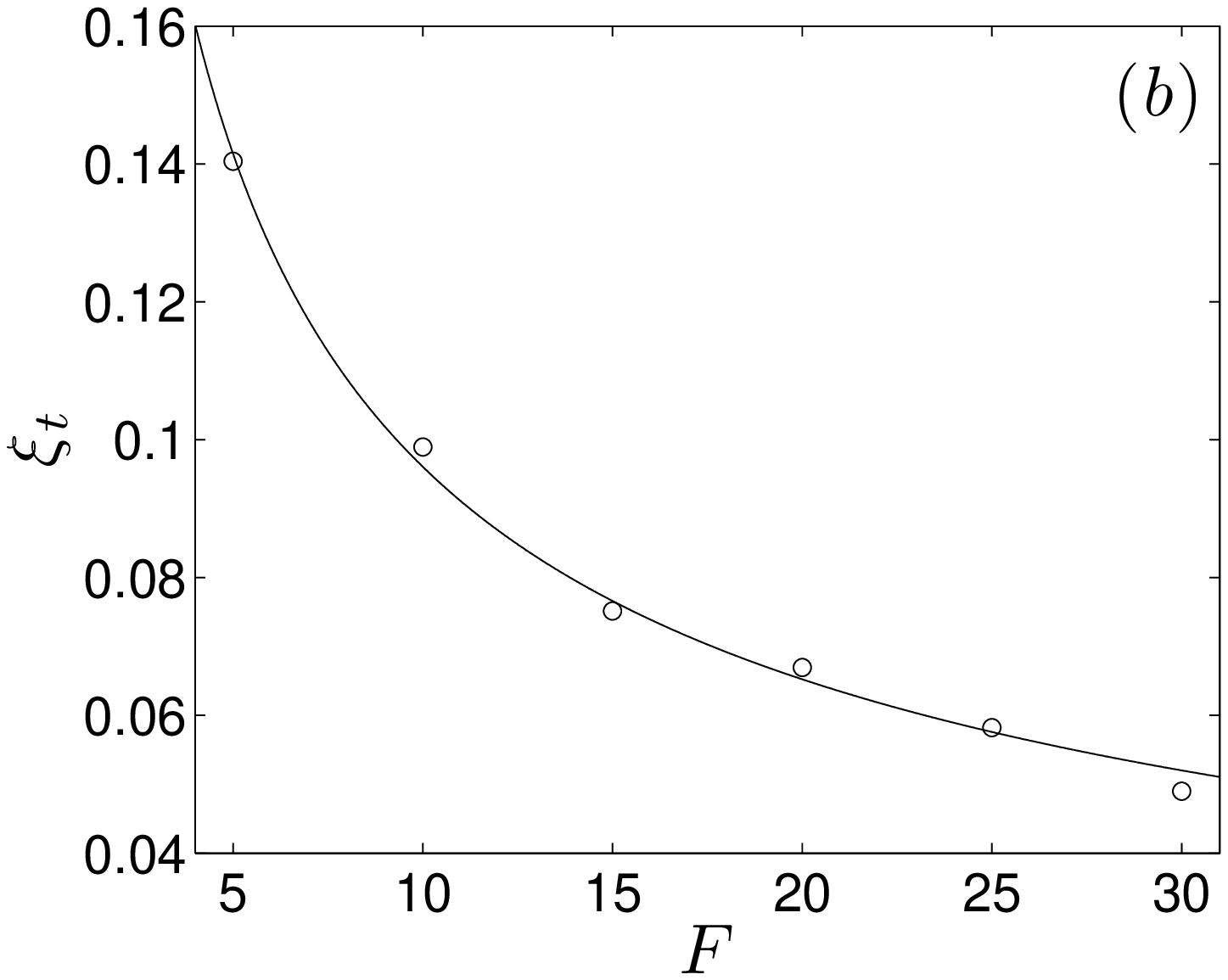}    
  \end{center}
\caption{The variation of two characteristic time scales with external forcing $F$. 
(a)~The variation of the normalized inverse of the leading order Lyapunov exponent 
$\lambda_1^{-1}/N$. Data are shown for 5 different system sizes as indicated in the 
legend. The solid line is a curve fit through the data of the form 
$\lambda_1^{-1}/N = 0.007 + 5.9 F^{-2.6}$. (b)~The variation of the average 
correlation time $\xi_t$ with $F$ for a system size of $N=35$. The symbols are results 
from simulation and the solid line is the power-law $\xi_t = 0.35 F^{-0.56}$ .}
\label{fig:timeF}
\end{figure}

We now compare several measures of characteristic time and length scales for  
increasing values of the magnitude of external forcing. Both the temporal 
and spatial scales decrease with increasing forcing as expected. The variation of 
these scales are well described by a power-law of the functional form given by 
Eq.~(\ref{eq:powerlaw}).  The variation of the inverse leading order Lyapunov 
exponent and the correlation time are shown in Fig~\ref{fig:timeF}.  The 
symbols are results from our simulations for each value of $F$. The value  
of $\lambda_1^{-1}$ yields a time scale related to the predictability of the 
system. The solid line in Fig~\ref{fig:timeF}(a) is a power-law curve fit using 
the values shown in Table~\ref{table:powerlaws}. The inverse Lyapunov exponents 
have been scaled by the system size $N$ so that all data can be represented on a single plot. 
The value of $a = 0.15$ represents $\lambda_1^{-1}/N$ for infinite $F$.

Figure~\ref{fig:timeF}(b) illustrates the variation of the average correlation time $\xi_t$ with 
increased forcing and over the 10 different initial conditions for a large system 
with $N=35$. The solid line is the power-law curve fit using the values of 
Table~\ref{table:powerlaws}. The fluctuations of the correlation time about 
the mean value over the different initial conditions is quite noisy and yields a 
coefficient of variation of 7.7\%. In the limit of infinite forcing the average 
correlation time $\xi_t \approx 0$.  Our results suggest that the predictability 
decreases faster than the correlation time indicating the possibility of at 
least two different time scales.
\begin{table}[tbp]
\begin{center}
\begin{tabular}
[c]{l@{\hspace{1.5cm}}l@{\hspace{1.5cm}}l@{\hspace{1.5cm}}l}
 & $a$ & $b$ & $\alpha$ \\ \hline \hline
$\delta_\lambda$       & 0.93     & - 4.03 & 1.32 \\ \hline
$\lambda_1^{-1}/N$       & 0.007    & 5.9 & 2.6 \\ 
$\xi_t$                         & $0$ & 0.35 & 0.56 \\ \hline 
$\xi_\delta$                 & 1.12 & 39.74 & 2.2 \\ 
$\xi_L$ & 4.38 & -290.8 & 3.5 \\ \hline
\end{tabular}
\end{center}
\caption{The coefficients describing power-law trends in the data 
from numerical simulations exploring the chaotic patterns as the 
magnitude of the external forcing is varied. The first row is for the 
dimension density $\delta_\lambda$, the next two rows describe temporal 
scales given by the inverse leading order Lyapunov exponent $\lambda_1^{-1}$ 
and correlation time $\xi_t$, and the final two rows are spatial scales given by 
the natural chaotic length scale $\xi_\delta$ and the pattern wavelength $\xi_L$.  The 
functional form of the power-law used is $a + b F^{-\alpha}$. }
\label{table:powerlaws}
\end{table}

\begin{figure}[htb]
  \begin{center}
    \includegraphics[width=0.45\textwidth]{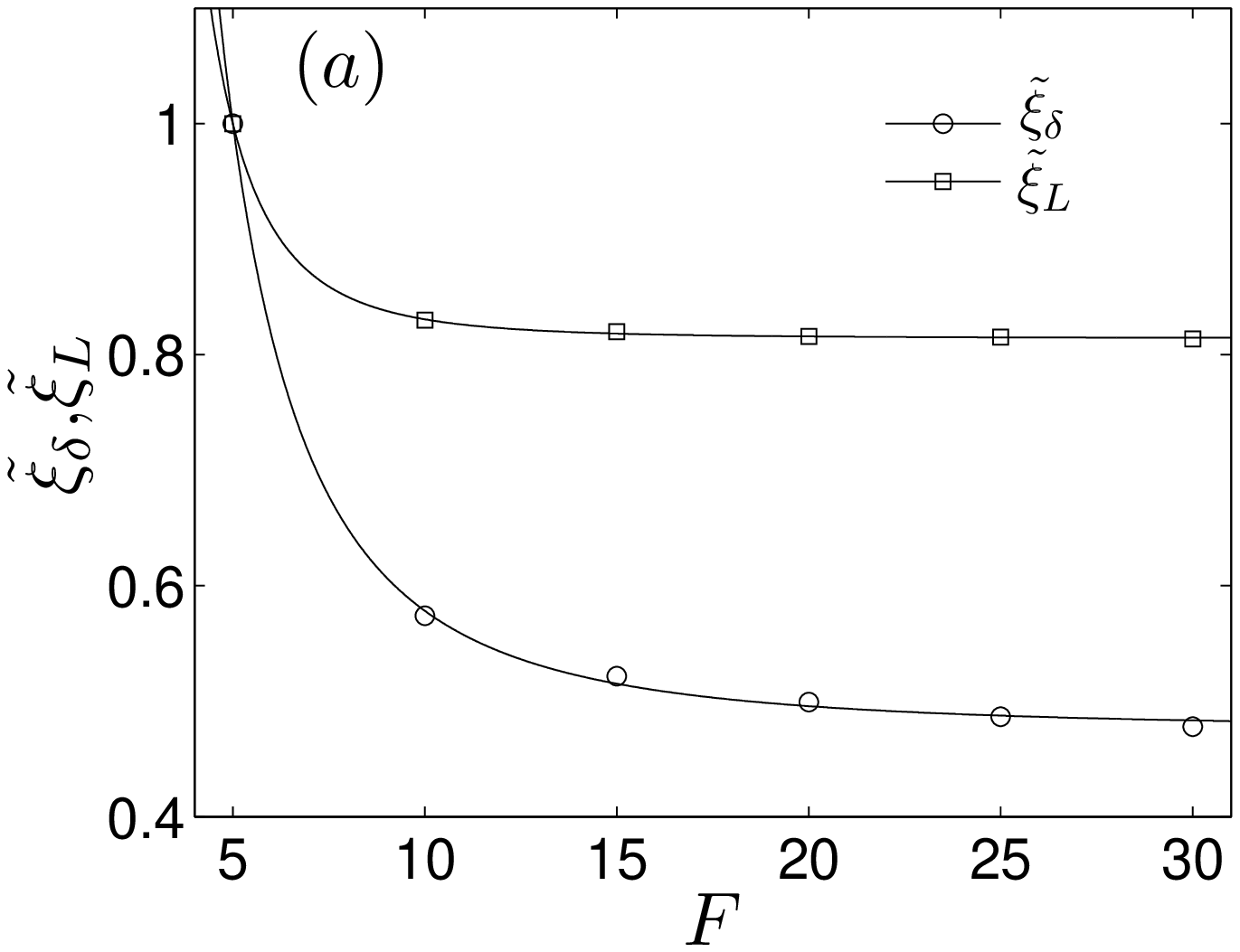}
    \includegraphics[width=0.45\textwidth]{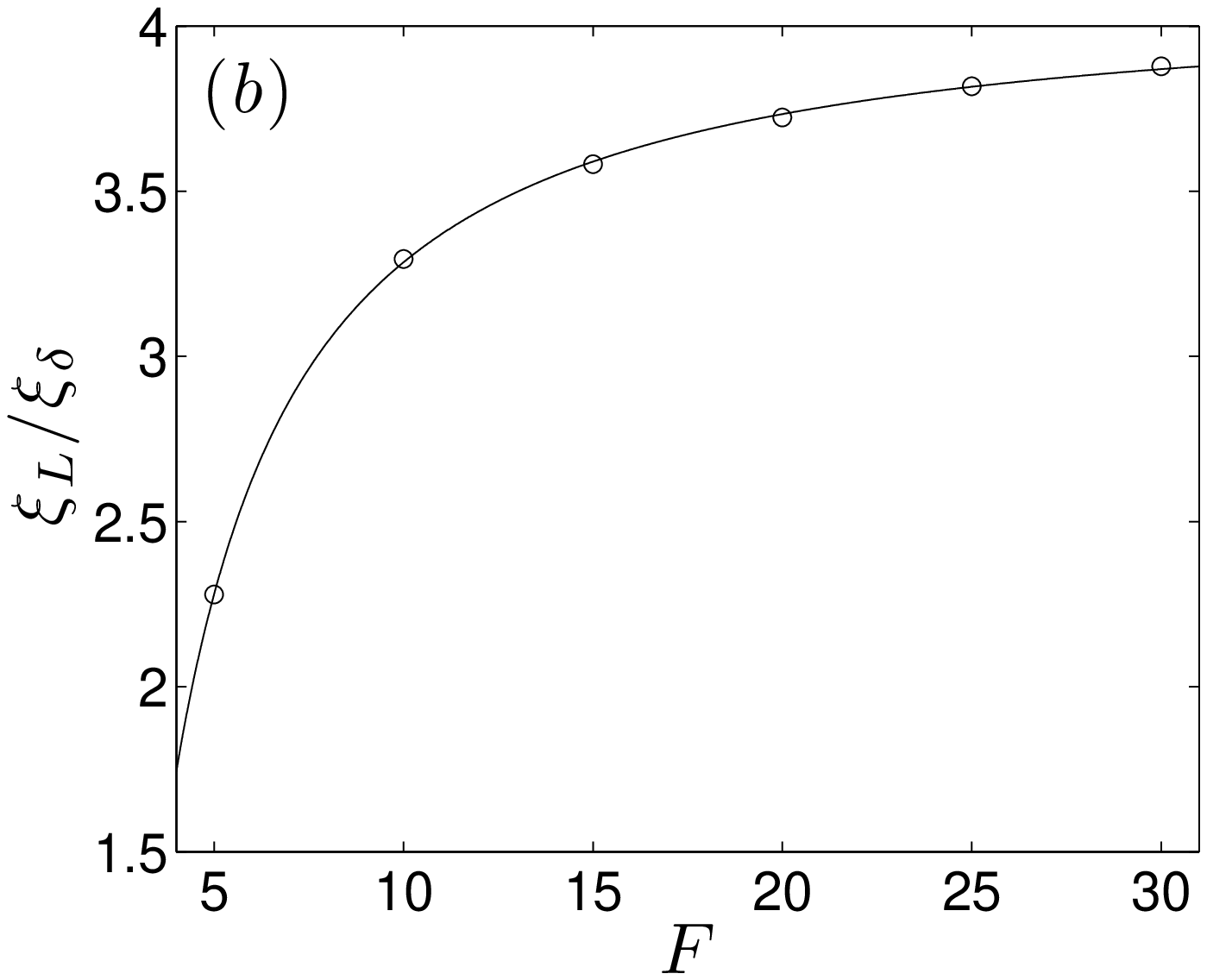}    
  \end{center}
\caption{The variation of two characteristic spatial scales with external forcing $F$ 
for a system size of $N=35$.  (a)~The normalized natural chaotic length scale $\tilde{\xi}_\delta$ 
(circles) and the normalized pattern wavelength $\tilde{\xi}_L$ (squares). The 
normalization factors used are the respective values of $\xi_\delta$ and $\xi_L$ 
at $F=5$.  The power-law curve fits for the unscaled length scales are 
$\xi_\delta = 1.12 + 39.74 F^{-2.2}$ and $\xi_L = 4.38 - 290.8 F^{-3.5}$. (b) 
The ratio $\xi_L/\xi_\delta$ yields an estimate for the number of chaotic degrees 
of freedom per pattern wavelength.}
\label{fig:spaceF}
\end{figure}

The pattern wavelength and the chaotic length scale both decrease with a 
power-law variation given by Eq.~(\ref{eq:powerlaw}) for increasing values 
of the magnitude of the external forcing.  The coefficients of the power-law 
variation are given in Table~\ref{table:powerlaws}.  The variation of these 
length scales is shown in Fig.~\ref{fig:spaceF}(a). The pattern wavelength 
is represented using square symbols and the natural chaotic length scale 
is represented using circle symbols.  The length scales have been normalized 
by their magnitude at $F=5$ to facilitate a comparison using a single plot. The 
normalized length scales are referred to as $\tilde{\xi}_L$ and $\tilde{\xi}_0$, 
respectively. The coefficient of variation, over the different initial conditions, for the chaotic 
length scale is quite small $\sim 0.01 \%$ whereas the pattern wavelength 
measurements are quite noisy with a coefficient of variation of $\sim 31\%$.  

It is clear that the natural chaotic length scale decreases more rapidly than the 
pattern wavelength.  The ratio of these two length scales yields an estimate for 
the number of chaotic degrees of freedom per pattern wavelength and is shown in 
Fig.~\ref{fig:spaceF}.  For small values of the forcing there are 
approximately 2 degrees of freedom per wavelength which increases to 
nearly 4 degrees of freedom per wavelength for large values of the forcing. 
The separation of these two length scales suggests that for larger values 
of the external forcing significant contributions to the overall disorder are 
from sub-wavelength structures in the pattern dynamics.  A clear signature 
of the sub-wavelength chaotic degree of freedom was not found using 
our spatial diagnostics.

\section{Conclusions}

Using a phenomenological model relevant to fluid convection and the atmosphere we 
have shown that the variation of the fractal dimension with system parameters 
can provide physical insights into fundamental features of high-dimensional chaos. We 
have used the fractal dimension to provide an approximate value of the number of 
chaotic degrees of freedom in the system and to estimate length scales describing 
the underlying chaotic dynamics.

If one considers only the space-time diagnostics of the correlation time 
and pattern wavelength the results exhibit significant fluctuations despite the use of 
very long-time simulations for numerous initial conditions.  This is in contrast to what 
is found when using the fractal dimension and suggests the possibility of features  
of the dynamics that have yet to be quantified in detail.  Our results indicate that the 
length scale describing the deviations from extensivity is approximately equal to 
the natural chaotic length scale which suggests the possibility of important spatial 
structures that contribute significantly to the chaotic dynamics.  Clearly identifying 
such features, and relating them to experimentally accessible quantities, remains an important 
open challenge in the characterization of systems driven far-from-equilibrium.
 
For systems of small size with low values of external forcing we find very complicated dynamics 
with windows of periodic and chaotic dynamics. This is similar to what has been found both 
numerically~\cite{paul:2001} and experimentally~\cite{bodenschatz:2000} in fluid systems 
such as Rayleigh-B\'{e}nard convection. For intermediate forcing, extensive chaos emerges 
with the important feature of significant deviations from extensivity for incremental 
changes in system size. Our results suggest that such behavior may be present in 
experimentally accessible fluid systems.  The ratio of our measured length scales, such 
as the ratio of the chaotic length scale $\xi_\delta$ to the wavelength of the deviations from 
extensivity $\xi_0$ do not yield integer values as found for the 1D complex Ginzburg-Landau 
equation~\cite{fishman:2006}. This is perhaps also due in part to the discrete spatial 
nature of the Lorenz-96 model.

The variation of the fractal dimension with external forcing yields insights into 
the manner in which the chaotic length scale decreases as 
the strong driving limit is approached. In the case of increasing system size while 
holding the forcing fixed, extensive chaos occurs when the 
exponent relating the dimension and system size is equal to the number of spatially 
extended directions. However, a similar theoretical understanding 
of the growth in the fractal dimension with increased forcing remains an 
open challenge.  The fact that our results are independent of system size is  
promising and suggests that perhaps this is an underlying feature of some 
generality.  It is anticipated that our results will be useful in guiding future 
efforts to explore the extensive chaos of experimentally accessible systems.

\bigskip

\noindent Acknowledgments: The computations were conducted using the 
resources of the Advanced Research Computing center at Virginia Tech and 
the research was supported by NSF grant no. CBET-0747727.

\end{document}